\documentclass[twocolumn]{aastex63}
\usepackage{natbib}
\usepackage{empheq}
\usepackage{epstopdf}
\usepackage{subfigure}
\usepackage{amsmath,amsfonts,amssymb}
\usepackage{multirow}
\usepackage{graphicx}
\begin{document}

\title{Black Hole Mass Function of Coalescing Neutron Star-Black Hole Binary Systems: The Prospect of Reconstruction with the Gravitational Wave Observations}
\author{Shao-Peng Tang}
\affil{Key Laboratory of Dark Matter and Space Astronomy, Purple Mountain Observatory, Chinese Academy of Sciences, Nanjing, 210033, People's Republic of China}
\affil{School of Astronomy and Space Science, University of Science and Technology of China, Hefei, Anhui 230026, People's Republic of China}
\author{Hao Wang}
\affil{Department of Physics and Astronomy, Purdue University, 525 Northwestern Avenue, West Lafayette, IN 47906, USA}
\author{Yuan-Zhu Wang}
\affil{Key Laboratory of Dark Matter and Space Astronomy, Purple Mountain Observatory, Chinese Academy of Sciences, Nanjing, 210033, People's Republic of China}
\author{Ming-Zhe Han}
\affil{Key Laboratory of Dark Matter and Space Astronomy, Purple Mountain Observatory, Chinese Academy of Sciences, Nanjing, 210033, People's Republic of China}
\affil{School of Astronomy and Space Science, University of Science and Technology of China, Hefei, Anhui 230026, People's Republic of China}
\author{Yi-Zhong Fan}
\affil{Key Laboratory of Dark Matter and Space Astronomy, Purple Mountain Observatory, Chinese Academy of Sciences, Nanjing, 210033, People's Republic of China}
\affil{School of Astronomy and Space Science, University of Science and Technology of China, Hefei, Anhui 230026, People's Republic of China}
\author{Da-Ming Wei}
\affil{Key Laboratory of Dark Matter and Space Astronomy, Purple Mountain Observatory, Chinese Academy of Sciences, Nanjing, 210033, People's Republic of China}
\affil{School of Astronomy and Space Science, University of Science and Technology of China, Hefei, Anhui 230026, People's Republic of China}
\email{Corresponding author: yzfan@pmo.ac.cn (YZF)}

\begin{abstract}
The discovery of gravitational waves from compact objects coalescence opens a brand-new window to observe the universe. With more events being detected in the future, statistical examinations would be essential to better understand the underlying astrophysical processes. In this work we investigate the prospect of measuring the mass function of black holes that are merging with the neutron stars. Applying Bayesian parameter estimation for hundreds of simulated neutron star\textendash black hole (NSBH) mergers, we find that the parameters for most of the injected events can be well recovered. We also take a Bayesian hierarchical model to reconstruct the population properties of the masses of black holes, in the presence of a low mass gap, both the mass gap and power-law index ($\alpha$) of black hole mass function can be well measured, thus we can reveal where the $\alpha$ is different for binary black hole (BBH) and NSBH systems. In the absence of a low mass gap, the gravitational wave data as well as the electromagnetic data can be used to pin down the nature of the merger event and then measure the mass of these very light black holes. However, as a result of the misclassification of BBH into NSBH, the measurement of $\alpha$ is more challenging and further dedicated efforts are needed.
\end{abstract}


\section{Introduction} \label{sec:intro}
The successful detection of a gravitational wave (GW) signal from the merger of a binary black hole (BBH) by Advanced LIGO \citep[aLIGO;][]{2016PhRvL.116f1102A}) on 2015 September 14 marks the onset of the era of GW astronomy, which opens a new window into observing the universe. Since then, dozens of GW events have been detected \citep{2019PhRvX...9c1040A}, including 10 confident detections of BBH mergers, a binary neutron star (BNS) merger event GW170817 \citep{2017PhRvL.119p1101A} with associated gamma-ray burst \citep{2017ApJ...848L..14G} and macronova/kilonova \citep{2017ApJ...848L..12A,2017Natur.551...67P}, and candidates with low false alarm rates (FAR) claimed at the LIGO/Virgo O3 public alerts. In a few years, aLIGO and Advanced Virgo (AdV) are anticipated to reach their design sensitivities, therefore many more GW signals will be detected \citep{2018LRR....21....3A}. Coalescing BNS and neutron star\textendash black hole (NSBH) binaries attract wide attention, because, in addition to giving rise to GWs, these mergers can also produce electromagnetic transients such as short/long-short GRBs and macronovae/kilonovae, as widely investigated in the literature \citep[e.g.,][]{1989Natur.340..126E,1998ApJ...507L..59L}. In the absence of GW observations, the identification of macronova/kilonova signals in the afterglow of a few short/long-short GRBs provides the strongest support to their compact object merger origin \citep[see][and the references therein]{2016NatCo...712898J}. The GW/GRB/macronova association provides a wealth of physical information about the source(s) and allows novel tests of fundamental physics \citep[e.g.,][]{1999BASI...27..627S,2013PhRvL.111g1101D,2016ApJ...827...75L,2016GReGr..48...95M,2016PhRvD..94b4061W,2017CQGra..34h4002P}, as demonstrated in the case of GW170817/GRB 170817A/AT2017gfo \citep[e.g.,][]{2017ApJ...848L..13A,2017ApJ...851L..18W}. Moreover, with the increasing sensitivities of the LIGO/Virgo/KAGRA detectors, the number of events will accumulate considerably in the next few years, reliable statistical studies will become possible.

\begin{figure*}
\centering
\includegraphics[width=1.0\columnwidth]{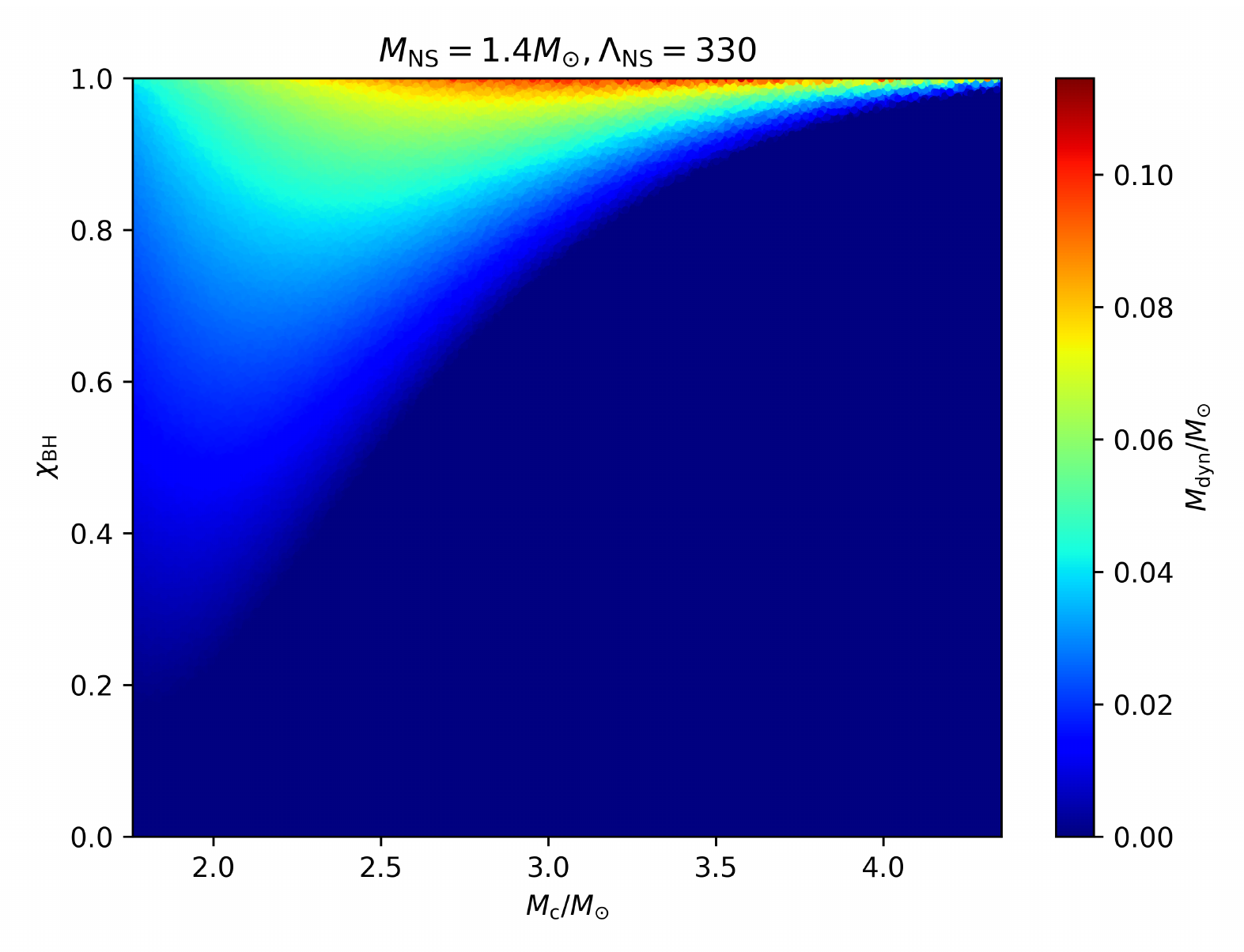}
\includegraphics[width=1.0\columnwidth]{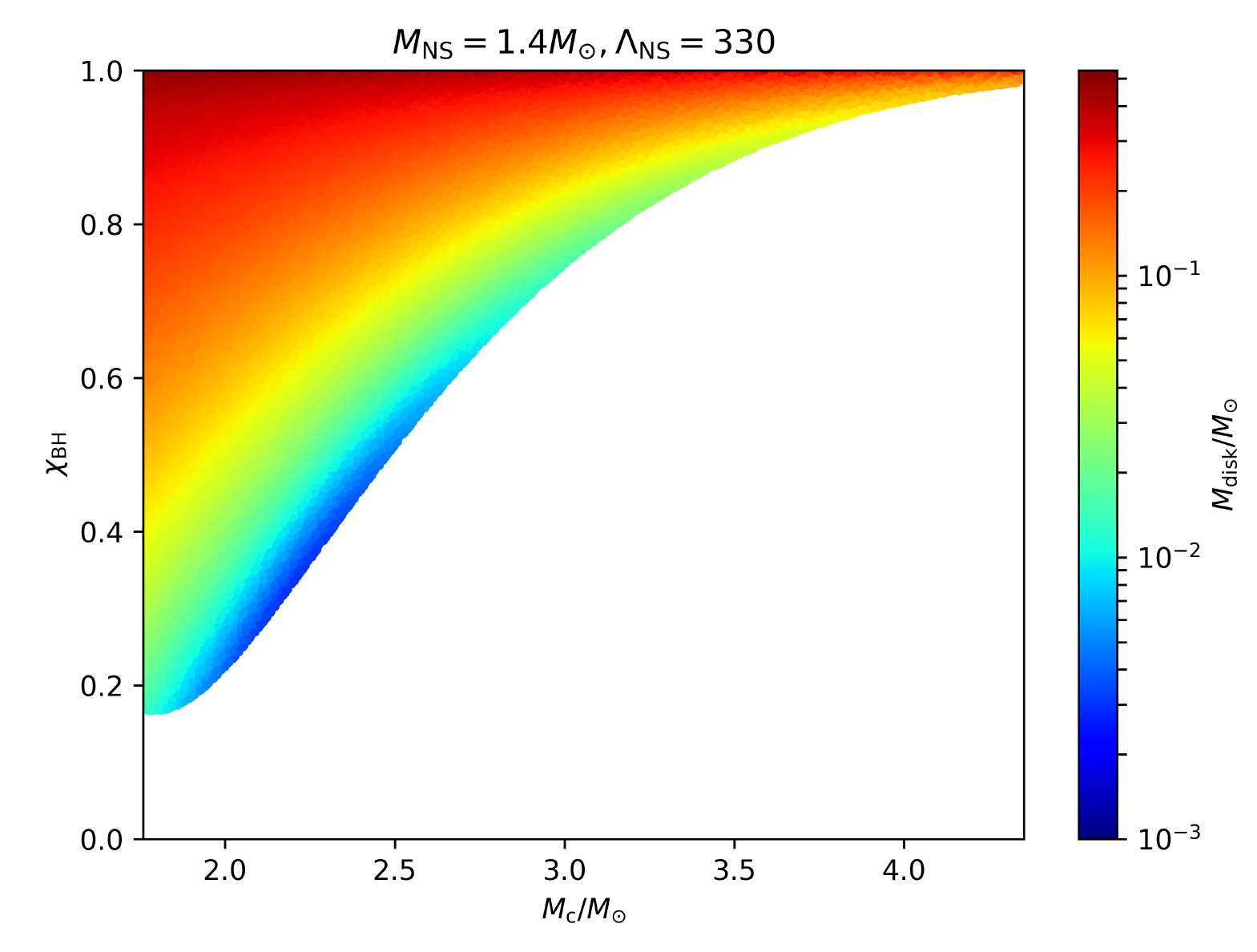}
\caption{Dynamical ejecta masses (left panel) and disk masses (right panel) produced by different chirp masses $\mathcal{M}_{\rm c}$ and the spins of BHs $\chi_{\rm BH}$.}
\label{fig:ejedisk}
\hfill
\end{figure*}
In this work, we focus on the black hole mass function (hereafter BHMF) of the merging NSBH binaries. Though NSBH binary systems have not been identified in the Galaxy yet, they are widely believed to exist in the universe \citep{2010CQGra..27q3001A} and the NSBH merger model for long-short GRB 060614 has been adopted to reproduce the luminous  macronova/kilonova signal \citep{2015ApJ...811L..22J,2015NatCo...6.7323Y}. Due to the current limited samples of stellar mass BHs, the BHMFs are not well determined, yet. However, previous studies have already identified some possible characteristics of BHMF from the observations of Galactic BHs. For example, the lightest black hole measured in X-ray binaries is $\sim 5 M_{\odot}$ \citep{2010ApJ...725.1918O}, much heavier than the upper limit of neutron stars. Such a result leads us to suspect the existence of a mass gap between the lightest black holes and the heaviest neutron stars. Population synthesis expects a high mass cutoff on the power-law mass distribution of black holes \citep{2015ApJ...806..263D}, because massive stars will lose their masses by stellar wind. Thanks to a high merger rate, such characteristics are expected to be identified in merging BBH systems via gravitational wave detection, as demonstrated in \citet{2017PhRvD..95j3010K}. For merging NSBH binary systems, the BHMF may be more challenging because of the expected smaller number of events. Nevertheless, an advantage of constructing BHMF of merging NSBH binaries is that the small chirp mass leads to better mass measurement \citep{1994PhRvD..49.2658C} for the same signal-to-noise ratio ({\rm S/N}). The other advantage is a good prior knowledge of neutron star distribution \citep{2013ApJ...778...66K}, which will compensate the large measurement error of mass ratio. One interesting question is whether the BHMFs are different between the merging NSBH and BBH binaries. This consideration is mainly motivated by the fact that neutron star distributions are slightly different in binary neutron stars and neutron star\textendash white dwarf binaries \citep{2012ApJ...757...55O,2013ApJ...778...66K}. On the other hand, BBHs may have multiple formation channels, such as binary stellar evolution and dynamical capture. While NSBH binaries have more difficulty forming through a dynamical process because of the small mass of neutron stars. The BHMF for different binary systems could then be different. However, it is beyond the scope of this work to quantify the prospect of identifying such a difference.

As for Bayesian parameter estimation with strain data of the NSBH merger events, the degeneracy between the mass and spin of BH may produce asymmetric errors or biases in mass measurements. Recently, \citet{2019A&A...625A.152B} showed that the electromagnetic (EM) counterparts information of the NSBH merger can help to break the degeneracies in the GW parameter space, leading to an unbiased estimation of BH mass compared to the sole GW data analysis \citep[Table IV]{2015PhRvD..91d2003V}. With the works of \citet{2016ApJ...825...52K} and \citet{2018PhRvD..98h1501F}, it is straightforward to use $(M_{\rm BH}, M_{\rm NS}, \chi_{\rm BH}, \Lambda_{\rm NS})$ to deduce the dynamical ejecta masses $M_{\rm dyn}$ and disk masses $M_{\rm disk}$ which are responsible for powering the electromagnetic emission. As shown in Fig.\ref{fig:ejedisk}, the chance of observing an NSBH merger with EM counterparts may be low for $M_{\rm BH}>5\,M_\odot$, due to the high ejecta mass requiring low mass and high spin for BH. Thus, for generality, we only consider the sole GW data injected in the Advanced LIGO/Virgo detectors with design sensitivities \citep{2018LRR....21....3A}. Therefore, we generate the simulated events and make a full Bayesian parameter estimation for the injected data, then apply a Bayesian hierarchical model to evaluate the prospect of characterizing BHMF.

Our work is organized as follows: in Section \ref{sec:methods} we introduce our BHMF model and the process of generating simulated events, analysis of single event using Bayesian parameter estimation, and Bayesian hierarchical model for constructing population properties of the BH masses. We present the results and discuss the implications in Section \ref{sec:results}. Section \ref{sec:discussion} contains our discussion and summary.

\section{Methods}\label{sec:methods}
For a long time, the mass function of stellar mass BHs has remained a topic of interest and a few models have been proposed/investigated \citep{2010ApJ...725.1918O,2015ApJ...806..263D,2017PhRvD..95j3010K,2019ApJ...882L..24A}. BBH merger events detected by Advanced LIGO/Virgo provide us with a powerful tool to measure the mass and spin of the source, which may trace the formation channels of BBH systems. However, the mass function of BH in NSBH systems still remains unknown because no such event has been reliably identified before. As reported in the LIGO/Virgo O3 public alerts (GraceDB\footnote{\url{https://gracedb.ligo.org/superevents/public/O3/}}), there were four NSBH candidates detected in the first six month run (S190814bv, S190910d, S190923y, and S190930t). This indicates a reasonably high merger rate of NSBH systems (note that the successful detection of NSBH events in late O2 or early O3 runs of Advanced LIGO/Virgo has been predicted by \citet{2017ApJ...844L..22L} based on the macronova/kilonova observations/modeling), thus it is possible to statistically reveal the BHMF with an accumulation of merger events in the next decade. This work aims to investigate the feasibility of reconstructing the BHMF with dozens of NSBH events.

\subsection{Injection Configurations}\label{sec:injection}
We use a phenomenological model to characterize the black hole and neutron star mass distributions and assume that they do not evolve with the redshift (given the limited distance range of the NSBH events detectable for Advanced LIGO/Virgo, this approximation is likely reasonable). The black hole population is assumed to obey a power-law distribution as adopted in \citet{2016PhRvX...6d1015A}. In addition, current observations in X-ray binaries suggest a cutoff at $\sim 5M_{\odot}$ \citep{2010ApJ...725.1918O}, while the population synthesis predicts cutoffs in both low and high mass bands \citep{2015ApJ...806..263D}. Very recently, a massive unseen companion with a mass of ${3.3}_{-0.7}^{+2.8}M_{\odot}$ was identified in the binary system \objectname{2MASS J05215658+4359220} \citep{2019Sci...366..637T}, and a few MassGap candidate events (S190924h, S190930s, and S191216ap) were claimed in GraceDB.
Thus, it is worth investigating both scenarios, i.e., the absence and presence of the low mass gap. In this work, we take the BHMF as,
\begin{gather}\label{eq:bhmf}
P(M_{\rm BH}) \propto M_{\rm BH}^{\rm -\alpha}{\rm exp}(-M_{\rm BH}/M_{\rm cut}),\nonumber \\
{\rm for} \ M_{\rm gap} \leqslant M_{\rm BH} \leqslant 95\,M_{\rm \odot},
\end{gather}
where we set the fiducial values to $\alpha=2.35$, $M_{\rm gap}=5\,M_{\rm \odot}$ ($3\,M_{\rm \odot}$; i.e., without the low mass gap), and $M_{\rm cut}=60\,M_{\rm \odot}$ following \citet{2016PhRvX...6d1015A} and \citet{2017PhRvD..95j3010K}. These parameters $\Lambda=\{\alpha, M_{\rm gap}, M_{\rm cut}\}$ are called hyperparameters that we try to reconstruct in Sec.\ref{sec:BHM}.

With the data of the first and second observing runs of Advanced LIGO/Virgo, \citet{2019ApJ...882L..24A} have further examined the BBH population properties (e.g., mass and spin distributions) with different phenomenological models. It is found that components of BBHs with large spins aligned with the orbital angular momentum are unlikely, while a low and restricted (LR) distribution of spin is favored. As show in Fig.\ref{fig:ejedisk}, the lack of GRB and kilonova observations for the four NSBH candidates also indicates that BHs may have a low spin or alternatively a too ``large" BH mass. Therefore, we adopt a low (\emph{L}) spin magnitude distribution with probability density function (PDF) $p(a_{\rm BH})=2\cdot(1-a_{\rm BH})$, and a restricted (\emph{R}) distribution of spin's tilt angle with PDF $p(\cos{\theta_{\rm BH}})=1\ (0<\cos{\theta_{\rm BH}}<1)$. For completeness, a flat (\emph{F}) spin magnitude distribution (uniformly spanning in range $[0,0.99]$) is also considered. Therefore, there will be four cases in our work, including
\begin{itemize}
\item Case A: With MassGap ($M_{\rm gap}=5M_{\odot}$), Low (\emph{L}) spin magnitude distribution;
\item Case B: With MassGap ($M_{\rm gap}=5M_{\odot}$), Flat (\emph{F}) spin magnitude distribution;
\item Case C: Without MassGap ($M_{\rm gap}=3M_{\odot}$), Low (\emph{L}) spin magnitude distribution;
\item Case D: Without MassGap ($M_{\rm gap}=3M_{\odot}$), Flat (\emph{F}) spin magnitude distribution.
\end{itemize}

\begin{table*}[]
\begin{ruledtabular}
\centering
\caption{Distributions for Injecting Signals and Priors of Bayesian Inference Adopted for GW Parameters $\vec{\theta}_{\rm GW}$}
\label{tb:distributions}
\begin{tabular}{cccc}
Names                                                                                      & Parameters                                                                     & Injection Configurations                              & Priors of Parameter Inference                   \\ \hline
Source frame mass of BH                                                        & $\frac{m_1/M_{\odot}}{(1+z)}$\tablenotemark{\tiny{a}}   & BHMF (Eq.(\ref{eq:bhmf}))                         & Bounded in (3, 100)                                  \\
Source frame mass of NS                                                        & $\frac{m_2/M_{\odot}}{(1+z)}$                                        & NSMF (Eq.(\ref{eq:nsmf}))                          & Bounded in (1.1, 2.1)                                \\
Detector frame chirp mass                                                       & $\mathcal{M}_{\rm c}/M_{\odot}$                                    & $\frac{(m_1m_2)^{3/5}}{(m_1+m_2)^{1/5}}$ & Uniform [1.5$\times$(1+z), 9.8$\times$(1+z)]    \\
Mass ratio                                                                                & $q$                                                                                  & $m_2/m_1$                                                & Uniform (0.011, 0.7)                                    \\
Spin magnitude of BH                                                              & $a_1$                                                                              & Low (\emph{L})/Flat (\emph{F})                & Uniform (0, 0.99)                                        \\
Spin magnitude of NS                                                              & $a_2$                                                                              & 0                                                                 & 0                                                                \\
Cosine of tilt angle between the BH's spin and $\vec{\rm L}$\tablenotemark{\tiny{b}} & $\cos{\theta_1}$                         & Restricted (\emph{R})                                 & Restricted (\emph{R})                               \\
Tilt angle between the NS's spin and $\vec{\rm L}$                & $\theta_2$                                                                       & 0                                                                 & 0                                                                \\
Azimuthal angle separating the spin vectors                           & $\phi_{12}$                                                                      & Uniform(0,$2\pi$)                                       & 0                                                                \\
Azimuthal position of $\vec{\rm L}$                                         & $\phi_{\rm JL}$                                                                & Uniform(0,$2\pi$)                                       & Uniform (0,$2\pi$)                                      \\
Luminosity distance                                                                 & $d_{\rm L}/{\rm Mpc}$                                                     & Uniform comoving-volume                         & Marginalized                                              \\
Inclination angle                                                                      & $\theta_{\rm JN}$                                                            & Uniform Sine                                              & Uniform Sine                                              \\
Right ascension                                                                      & ${\rm R.A.}$                                                                      & Uniform(0,$2\pi$)                                       & Uniform (0,$2\pi$)                                      \\
Declination                                                                              & ${\rm Decl.}$                                                                    & Uniform Cosine                                          & Uniform Cosine                                          \\
Coalescence phase                                                                & $\phi$                                                                               & 0                                                                 & Marginalized                                              \\
Polarization of GW                                                                 & $\Psi$                                                                               & Uniform(0,$\pi$)                                          & Uniform (0,$\pi$)                                         \\
Geocentric GPS time of the merger                                       & $t_{\rm c}/{\rm s}$                                                            & 60                                                                & Marginalized                                               \\
Tidal deformability of BH                                                        & $\Lambda_{1}$                                                                 & 0                                                                 & 0                                                                  \\
Tidal deformability of NS                                                        & $\Lambda_{2}$                                                                 & $\Lambda_{\rm NS}$                                 & 0                                                                  \\
\end{tabular}
\tablenotetext{\tiny{a}}{\leftline{$z$ is the cosmic redshift calculated with luminosity distance assuming ${\rm \Lambda CDM}$ cosmology}}
\tablenotetext{\tiny{b}}{\leftline{$\vec{\rm L}$ means the orbital angular momentum}}
\end{ruledtabular}
\end{table*}

As for the neutron star mass function (NSMF), a truncated gaussian distribution is adopted, i.e.,
\begin{gather}\label{eq:nsmf}
P(M_{\rm NS}) \! \propto \! \frac{1}{\sqrt{2\pi}\sigma}{\rm exp}\left[-\frac{(M_{\rm NS}-\mu)^2}{2\sigma^2}\right],\nonumber \\
{\rm for} \ M_{\rm min} \! \leqslant \! M_{\rm NS} \! \leqslant \! M_{\rm max},
\end{gather}
where $\mu$, $\sigma$, $M_{\rm min}=1.1\,M_{\rm \odot}$, and $M_{\rm max}=2.1\,M_{\rm \odot}$ are the mean value, standard deviation, lower and upper bounds of NS masses, respectively. Based on current observation data \citep{2013ApJ...778...66K}, we choose $\mu=1.33\,M_{\rm \odot}$ and $\sigma=0.12\,M_{\rm \odot}$ assuming that NS mass distribution in NSBH systems is similar to that in BNS. Though the minimum/maximum mass of NS is still uncertain, our choice of $M_{\rm NS} \in [1.1, 2.1]M_{\rm \odot}$ is reasonable \citep[e.g.,][]{2018MNRAS.481.3305S, 2019NatAs.tmp..439C,2020ApJ...888...45T}, and has little influence on our simulations due to the narrow distribution of NSMF (the probability of injecting very low/high NS mass is pretty low). Additionally, neutron stars would spin down due to the magnetic dipole radiation and lose their angular momentum during the long merging time scale. For simplicity, we only consider the nonrotating NS case, which is in agreement with expectations from Galactic BNS spin measurements \citep{2017ApJ...846..170T,2018PhRvD..98d3002Z}, and the approximation of fixing the spin of NS to zero when we inject signals has negligible effect on our study. \citet{2018PhRvL.121i1102D} showed that the relation between tidal deformability and mass of NS approximately obey $\Lambda(M) \propto M^{-6}$ in a relevant mass range. In this work we take $\Lambda_{1.4}=330$ \citep[e.g.,][]{2018PhRvL.121p1101A,2019ApJ...885...39J} and the tidal deformability is injected as $\Lambda_{\rm NS}(M_{\rm NS}) = 330\times{\left(M_{\rm NS}/1.4M_{\rm \odot}\right)}^{-6}$.

All parameters ($\vec{\theta}_{\rm GW}$) used to generate GW waveforms and their corresponding distributions are summarized in Table.\ref{tb:distributions}, where we take a uniform comoving-volume distribution up to $500 {\rm Mpc}$ for luminosity distance ($d_{\rm L}$), and a uniform sky distribution for the location parameters, i.e., right ascension (${\rm R.A.}$) and declination (${\rm decl.}$). To inject the simulated signals, an inspiral-only post-Newtonian waveform template named SpinTaylorT4Fourier is adopted, which is competent for components with arbitrary, precessing spins \citep{2014PhRvD..90l4029K,2015PhRvD..91d2003V}. Besides, we take the power spectral density (PSD\footnote{\url{https://dcc.ligo.org/LIGO-P1200087-v42/public}}) of design sensitivities into account, which is appropriate for NSBH merger in the aLIGO/AdV era \citep{2018LRR....21....3A}. We set a typical condition that the {\rm S/N} of a single interferometer satisfying ${\rm S/N} > 8.0$, as the definition of a GW event being ``detected.'' This approximately translates into a network ${\rm S/N}>12$, which is conventionally used as the threshold for a network GW detector to identify the GW signals \citep{2010CQGra..27q3001A,2017PhRvD..95j3010K,2019PASA...36...10T}.

\subsection{Single Event Analysis}\label{sec:inference}
To examine how well the parameters of BHMF can be constrained, we first perform a Bayesian parameter inference for each simulated event, using the $Bilby$ package \citep{2019ascl.soft01011A} and $PyMultinest$ sampler \citep{2016ascl.soft06005B}. Based on the Bayes' Theorem, the posterior PDF is proportional to the product of the prior PDF $p(\vec{\theta}_{\rm GW})$ and the likelihood $L(d^{\rm inj}|\vec{\theta}_{\rm GW})$ of the injected signal $d^{\rm inj}$ given the waveform model described by $\vec{\theta}_{\rm GW}$, i.e., $p(\vec{\theta}_{\rm GW}|d^{\rm inj}) \propto L(d^{\rm inj}|\vec{\theta}_{\rm GW})p(\vec{\theta}_{\rm GW})$. If we assume stationary Gaussian noise, then the log-likelihood of single detector usually takes the function form,
\begin{equation}
\label{eq:Likelihood}
{\rm log} L(\vec{\theta}_{\rm GW}) \! = \! -2 \int_{f_{\rm min}}^{f_{\rm max}} \frac{|d^{\rm inj}(f) \! - \! h(\vec{\theta}_{\rm GW},f)|^2}{S_{\rm n}(f)} df \! + \! C,
\end{equation}
where $S_{\rm n}(f)$, $d^{\rm inj}(f)$, and $h(\vec{\theta}_{\rm GW},f)$ represent the one-sided PSD of the noise, the injected signal, and the frequency domain waveform generated using parameter $\vec{\theta}_{\rm GW}$, respectively. Due to the lack of reliable numerical simulation based waveform template including tidal effect for NSBH merger, we only consider the frequencies bounded in the range of $(23\,{\rm Hz}, \, f_{\rm ISCO})$ to preview the situation of analyzing the future real data with an inspiral-only template, e.g., SpinTaylorT4Fourier. Furthermore, $f_{\rm ISCO}$ is calculated with the following formulae \citep{1972ApJ...178..347B,2019PhRvD.100h4031A},
\begin{equation}\label{eq:fisco}
\scriptsize
\begin{split}
Z_1 \! = \! 1 \! + \! (1 \! - \! \chi_{z}^2)^{1/3}[(1 \! + \! \chi_{z})^{1/3} \! + \! (1 \! - \! \chi_{z})^{1/3}], \quad Z_2 \! = \! (3\chi_{z}^2 \! + \! Z_1^2)^{1/2}, \\
f_{\rm ISCO} \! = \! \frac{6^{3/2}}{[3 \! + \! Z_2 \! - \! {\rm sign}(\chi_{z})\sqrt{(3 \! - \! Z_1)(3 \! +\! Z_1 \! + \! 2Z_2)}]^{3 \! / \! 2} \! + \! \chi_{z}}\frac{4400\,{\rm Hz}}{m_1^{\rm src} \! + \! m_2^{\rm src}},
\end{split}
\end{equation}
where $\chi_{z}=a_1\cos{\theta_1}$ is the projection of BH spin along the direction of orbital angular momentum, and $m_1^{\rm src}$, $m_2^{\rm src}$ are source frame masses of the components in unit of $M_{\rm \odot}$. This procedure has little influence on extracting the mass and spin information from GW signal, because the properties, e.g., chirp mass $\mathcal{M}_{\rm c}$, and mass ratio $q$, are predominantly determined by inspiral stage \citep{2012PhRvD..85l3007D}.

Though the real data recorded by Advanced LIGO/Virgo may suffer from glitch and nonstationary noise, which may produce biased PSD estimation. Some powerful tools, e.g., BayesLine and BayesWave, have been developed to solve this problem \citep{2015CQGra..32m5012C, 2015PhRvD..91h4034L,2019arXiv190811170T}. Here, we only consider the ideal case by assuming the PSD can be well estimated, and use the same PSD and waveform template as injecting signals to infer the GW parameters of each simulated event. The priors of $\vec{\theta}_{\rm GW}$ are listed in Table.\ref{tb:distributions}, where we marginalize the likelihood over the phase $\phi$, geocentric time $t_{\rm c}$, and luminosity distance $d_{\rm L}$ to accelerate Nest sampling \citep{2012PhRvD..85l2006A, 2018arXiv180510457L, 2019PASA...36...10T}. Due to the component masses $m_1$ and $m_2$ being partially degenerate, we sample the chirp mass $\mathcal{M}_{\rm c}$ and mass ratio $q$ instead of these parameters to improve the convergence rate of the stochastic sampler \citep{2019PhRvX...9a1001A}. Additionally, we request that the component masses are constrained in reasonable ranges (i.e., $m_1^{\rm src}\in[3,\,100]$, $m_2^{\rm src}\in[1.1,\,2.1]$) when we sample $\mathcal{M}_{\rm c}$ and $q$.

\subsection{Bayesian Hierarchical Model}\label{sec:BHM}
\begin{figure}
\centering
\includegraphics[width=1.0\columnwidth]{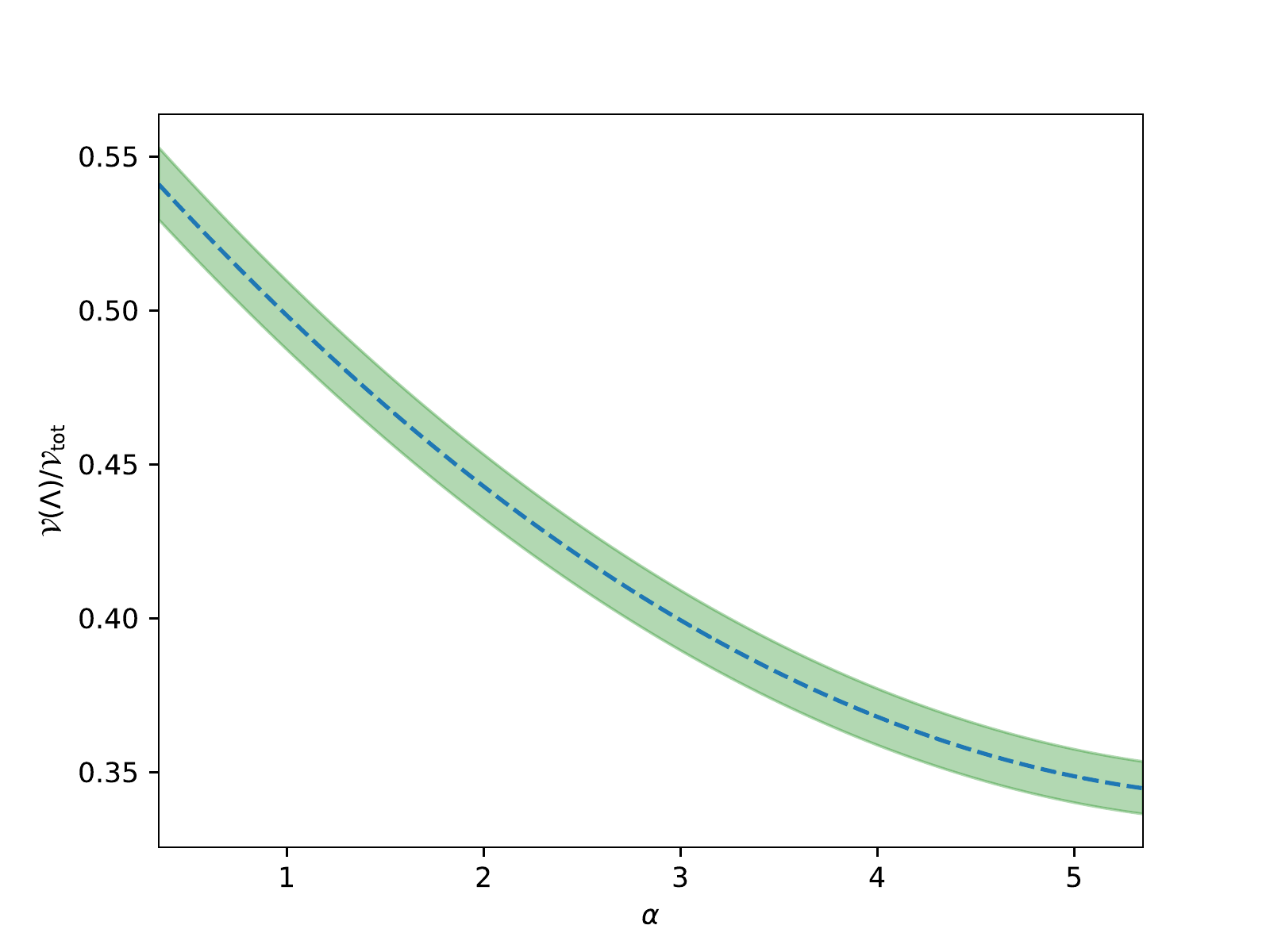}
\caption{Dashed line represents the relation between the ratio of visible volume to total spacetime volume ${\cal V}(\Lambda)/{\cal V}_{\rm tot}$ and hyperparameter $\alpha$, while the shaded area marks the $68\%$ confidence region.}
\label{fig:pdet}
\hfill
\end{figure}
Bayesian hierarchical inference \citep{2012PhRvD..86l4032A,2019PASA...36...10T} allows us to probe the population properties of an ensemble of events, and has been widely used in various fields, e.g., studying the evolutionary scenarios of binary stellar \citep{2018PhRvD..98h3017T}, revealing the origin of BHs from effective spin measurements \citep{2019JCAP...08..022F}, constructing mass distribution of galactic BNS \citep{2019ApJ...876...18F}, constraining of the equation of state (EoS) of NS \citep{2019PhRvD.100j3009H}, and investigating the jet properties of short gamma-ray bursts \citep[sGRB;][]{2019APS..APRH16006B}.

\begin{table*}
\scriptsize
\centering
\begin{ruledtabular}
\caption{The $5\%$-$95\%$ Confidence Intervals of Some Recovered Parameters}
\begin{tabular*}{1.0\textwidth}{cccccccc}
              & $\mathcal{M}_{\rm c}^{\rm src} (M_{\odot})$    & Mass ratio $q$    & $M_{\rm BH}^{\rm src} (M_{\odot})$    & $\chi_{z}$    & ${\rm R.A. (rad)}$    & ${\rm Decl. (rad)}$ & Network {\rm S/N}\\
\hline
Inferred & $2.5713_{-0.0040}^{+0.0044}$  & $0.1495_{-0.0444}^{+0.0547}$  & $8.2704_{-1.3584}^{+1.8861}$  & $0.2238_{-0.1152}^{+0.1057}$  & $-0.0816_{-0.0724}^{+0.0727}$  & $2.9660_{-0.0490}^{+0.0463}$  &\dots \\
Injected & $2.5732$     & $0.1500$     & $8.2599$     & $0.2489$     & $-0.1593$     & $3.0205$    & $18.3$ \\
\hline
Inferred & $2.4957_{-0.0019}^{+0.0020}$  & $0.5706_{-0.2061}^{+0.1154}$  & $3.8248_{-0.3518}^{+1.0384}$  & $0.3727_{-0.0495}^{+0.0476}$  & $0.7803_{-0.0466}^{+0.0519}$  & $2.5739_{-0.0793}^{+0.1093}$  &\dots \\
Injected & $2.5021$     & $0.2080$     & $6.6666$     & $0.4676$     & $0.7654$     & $2.5849$    & $13.2$ \\
\end{tabular*}
\label{tb:single}
\end{ruledtabular}
\end{table*}

\begin{figure*}
\centering
\includegraphics[width=0.48\columnwidth]{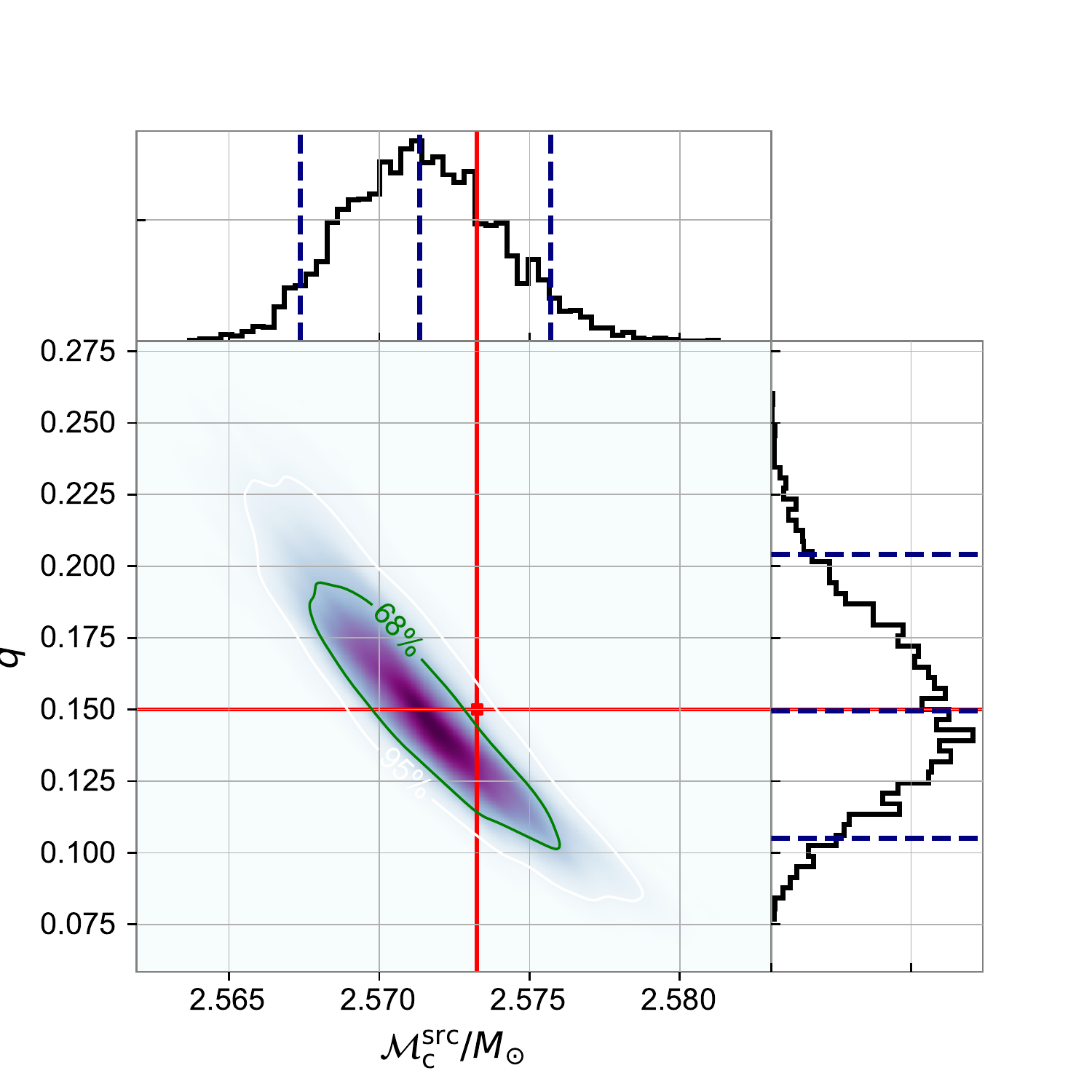}
\includegraphics[width=0.48\columnwidth]{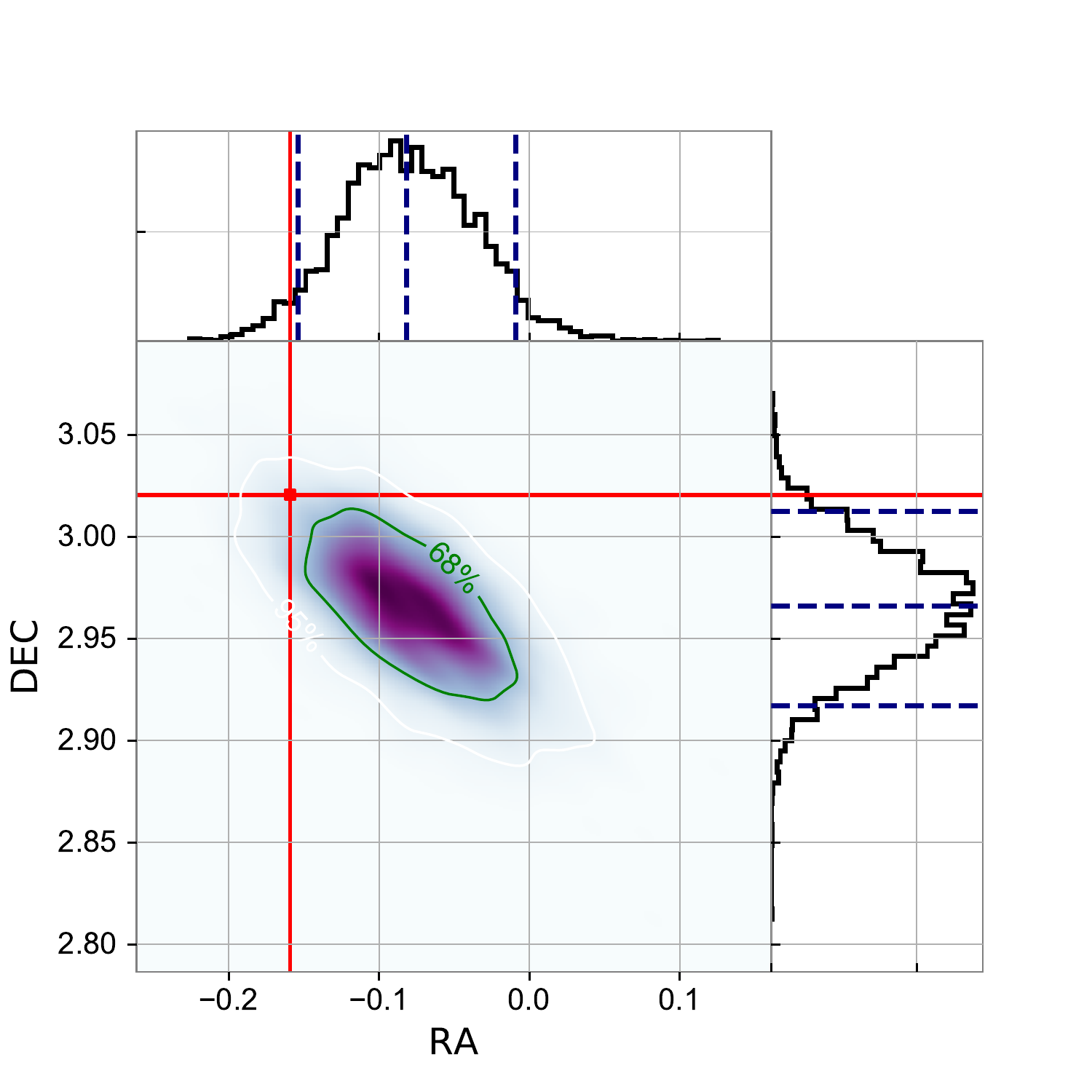}
\includegraphics[width=0.48\columnwidth]{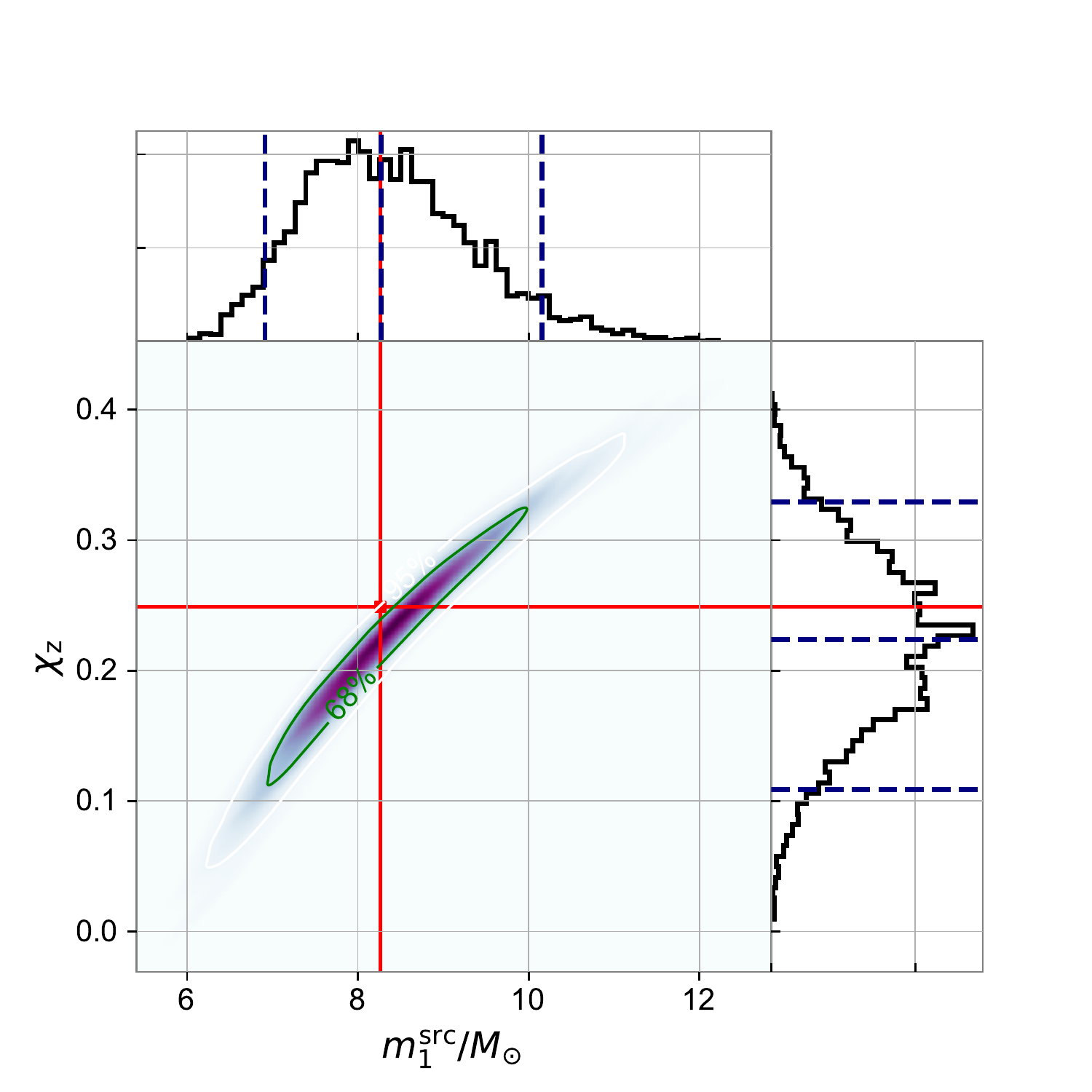}
\includegraphics[width=0.48\columnwidth]{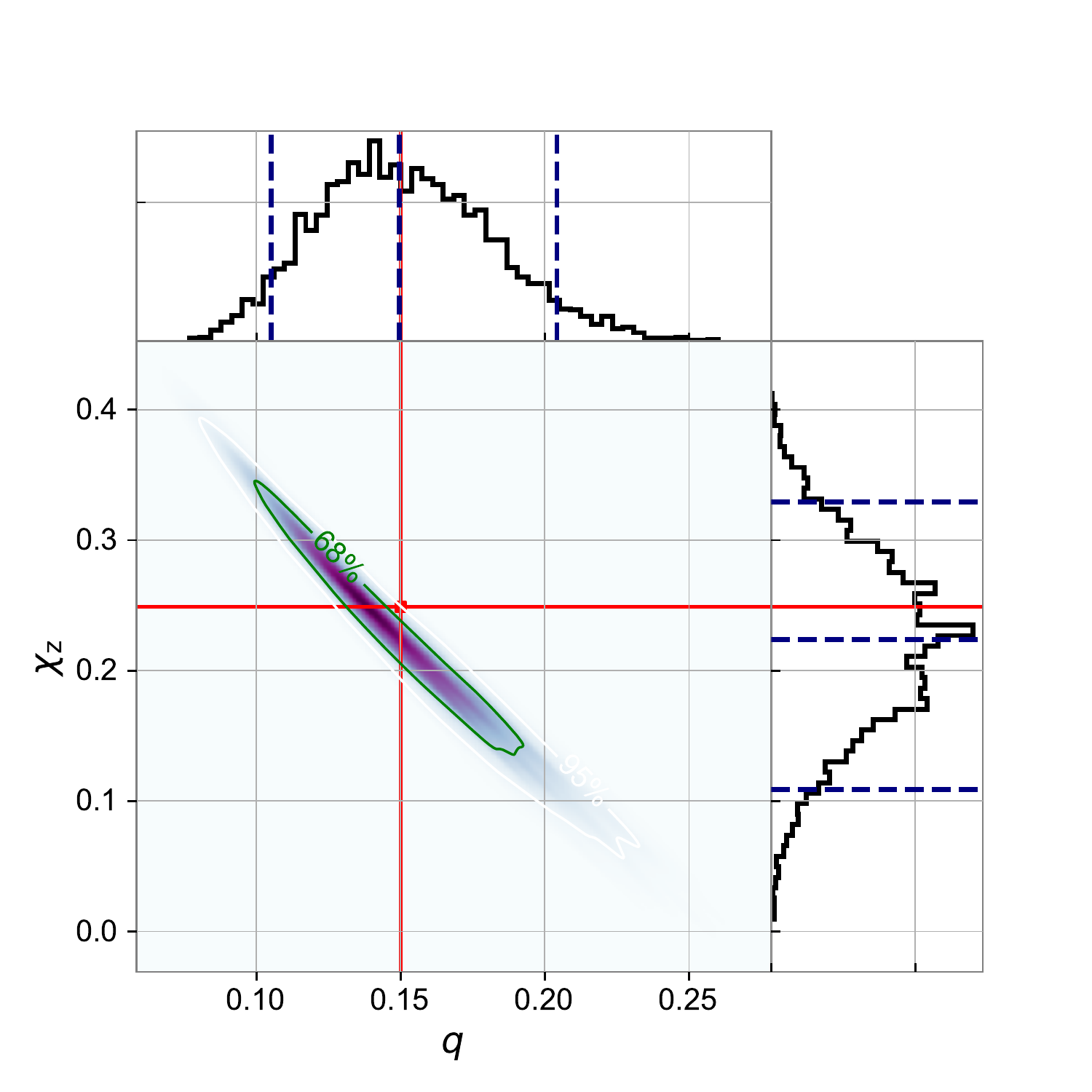}
\includegraphics[width=0.48\columnwidth]{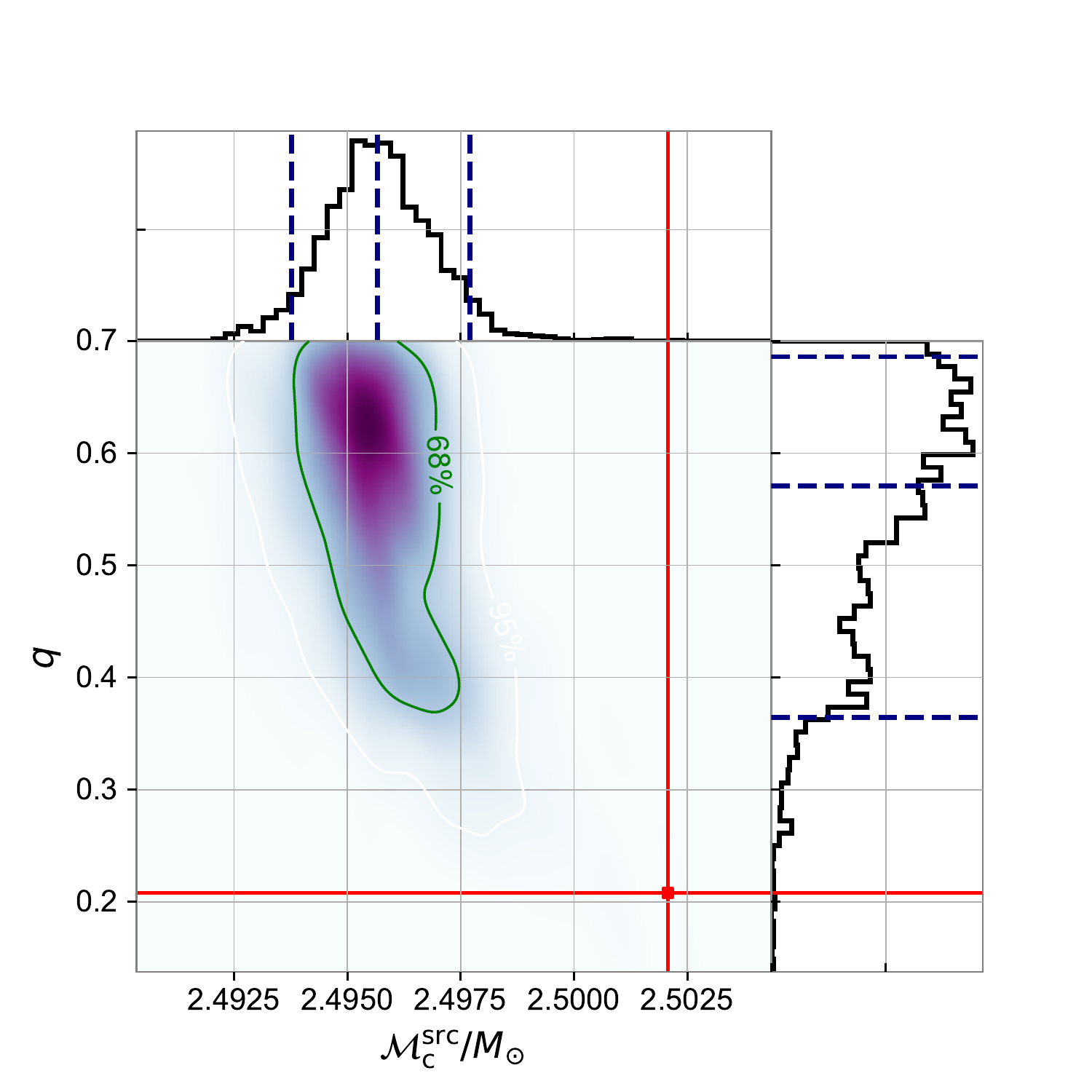}
\includegraphics[width=0.48\columnwidth]{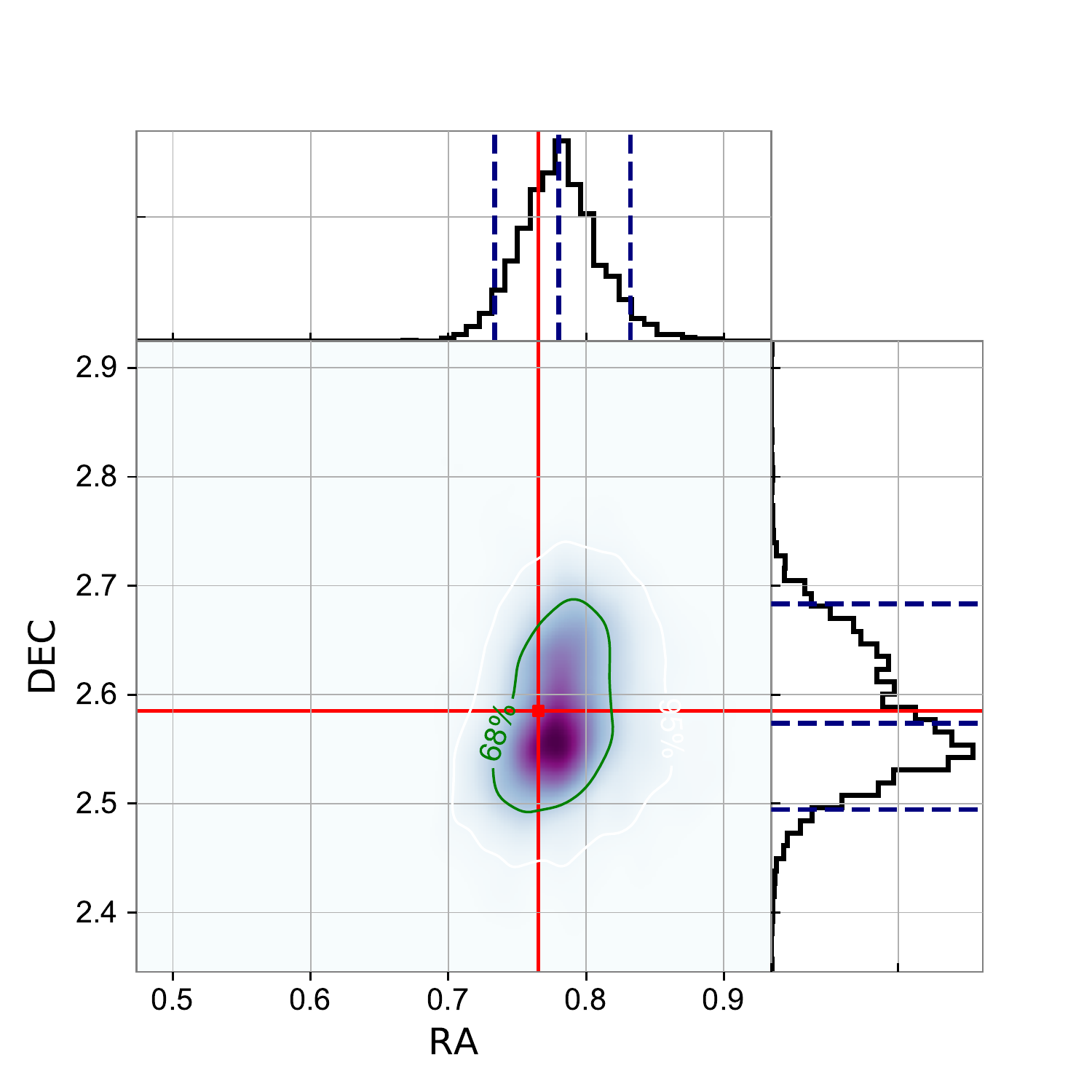}
\includegraphics[width=0.48\columnwidth]{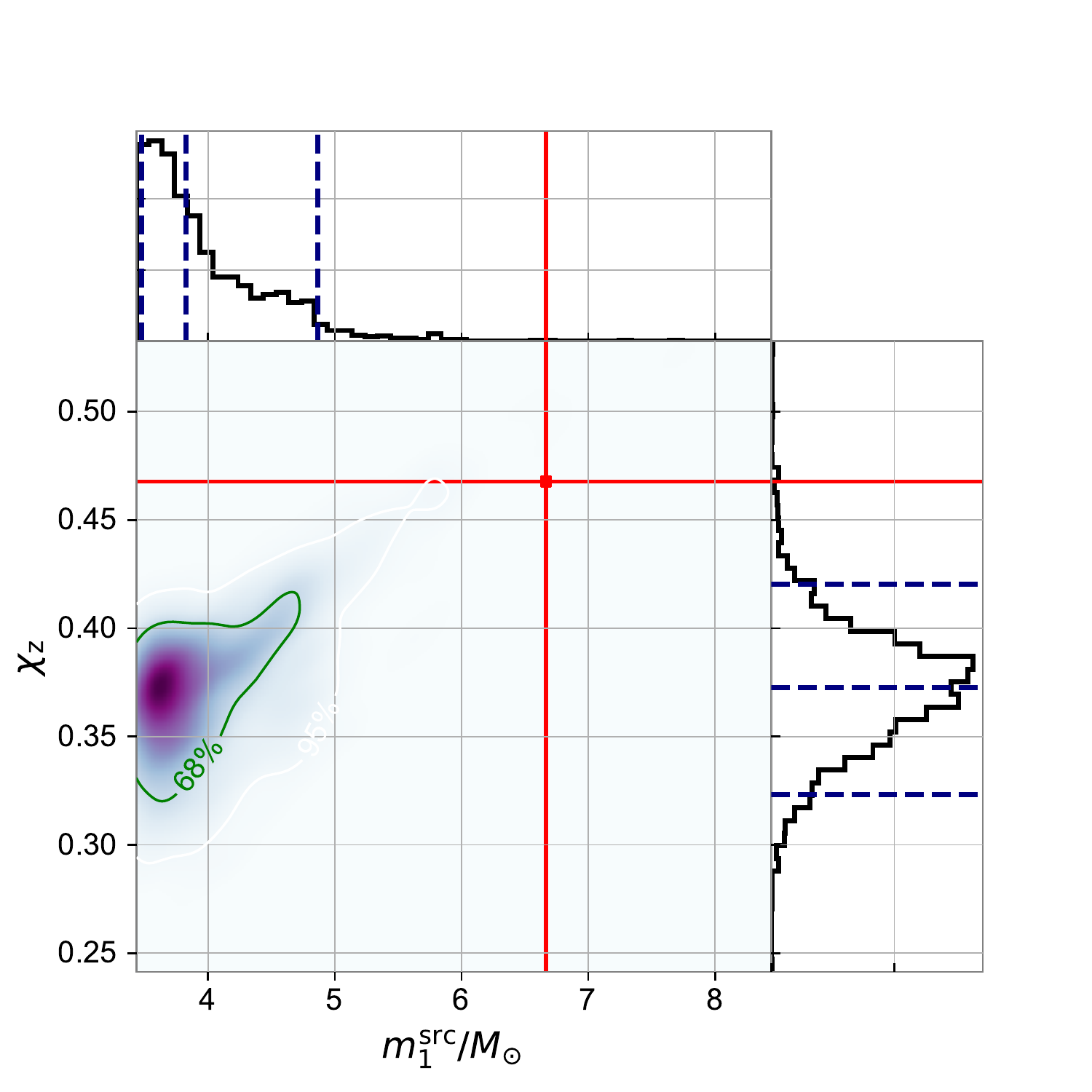}
\includegraphics[width=0.48\columnwidth]{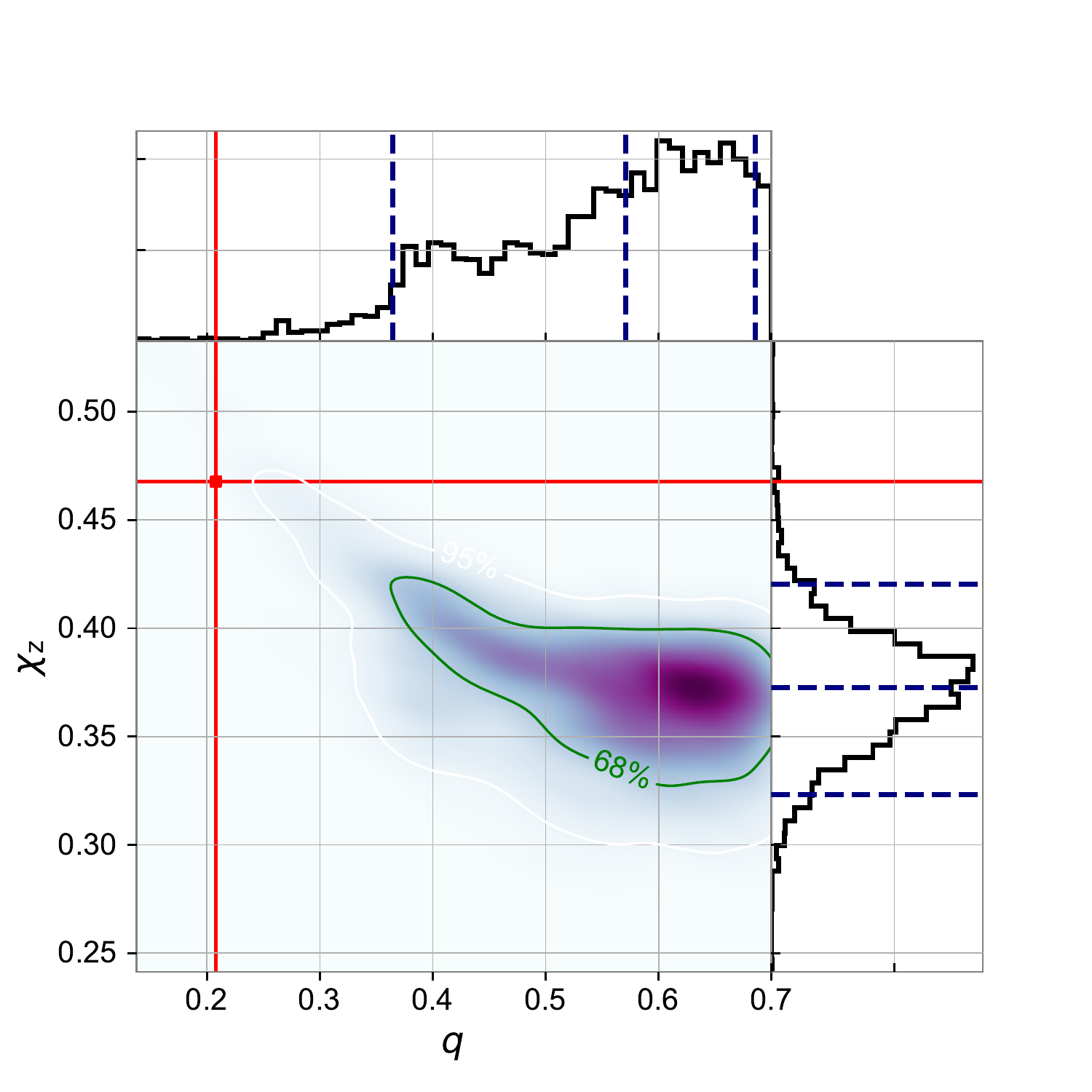}
\caption{The results of single event analysis. Red lines are injected values, and blue dashed lines represent ($5\%$, $50\%$, and $95\%$) percentiles.}
\label{fig:single}
\hfill
\end{figure*}

Based on the method introduced by \citet{2019PASA...36...10T} and \citet{2019arXiv191209708G}, we apply this technique to infer hyperparameters $\Lambda=\{\alpha, M_{\rm gap}, M_{\rm cut}\}$ with the likelihood
\begin{equation}\label{eq:bhm}
{\cal L}_{\rm tot}(\vec{d}, N | \Lambda) = \prod_{i}^N \frac{{\cal Z}_{\o}(d_{i})}{n_{i}} \sum_{k}^{n_{i}} \frac{\pi(\theta^{k}_{i}|\Lambda)}{\pi(\theta^{k}_{i}|{\o})},
\end{equation}
where $N$, $n_{i}$, ${\cal Z}_{\o}(d_{i})$, $\pi(\theta^{k}_{i}|\Lambda)$, and $\pi(\theta^{k}_{i}|{\o})$ represent the total number of events, the size of downsampled posterior samples, the Bayesian evidence of each event, the normalized BHMF, and the prior of the BH's source frame mass, respectively.

Through single event analysis described in Sec.\ref{sec:inference}, the Bayesian evidences are directly obtained by Nest sampling, and the samples of source frame masses of BHs ($m_1^{\rm src}$) can be transformed from the posterior samples of $\mathcal{M}_{\rm c}$, $q$, and the reconstructed $d_{\rm L}$ via
\begin{equation}
\label{eq:mcqtom12}
m_1^{\rm src}=\frac{q^{-3/5}(1+q)^{1/5}\mathcal{M}_{\rm c}}{1+z(d_{\rm L})},
\end{equation}
where $z$ is the cosmic redshift calculated with luminosity distance $d_{\rm L}$ assuming ${\rm \Lambda CDM}$ cosmology \citep{2016A&A...594A..13P}.

\begin{figure*}
\centering
\includegraphics[width=0.8\columnwidth]{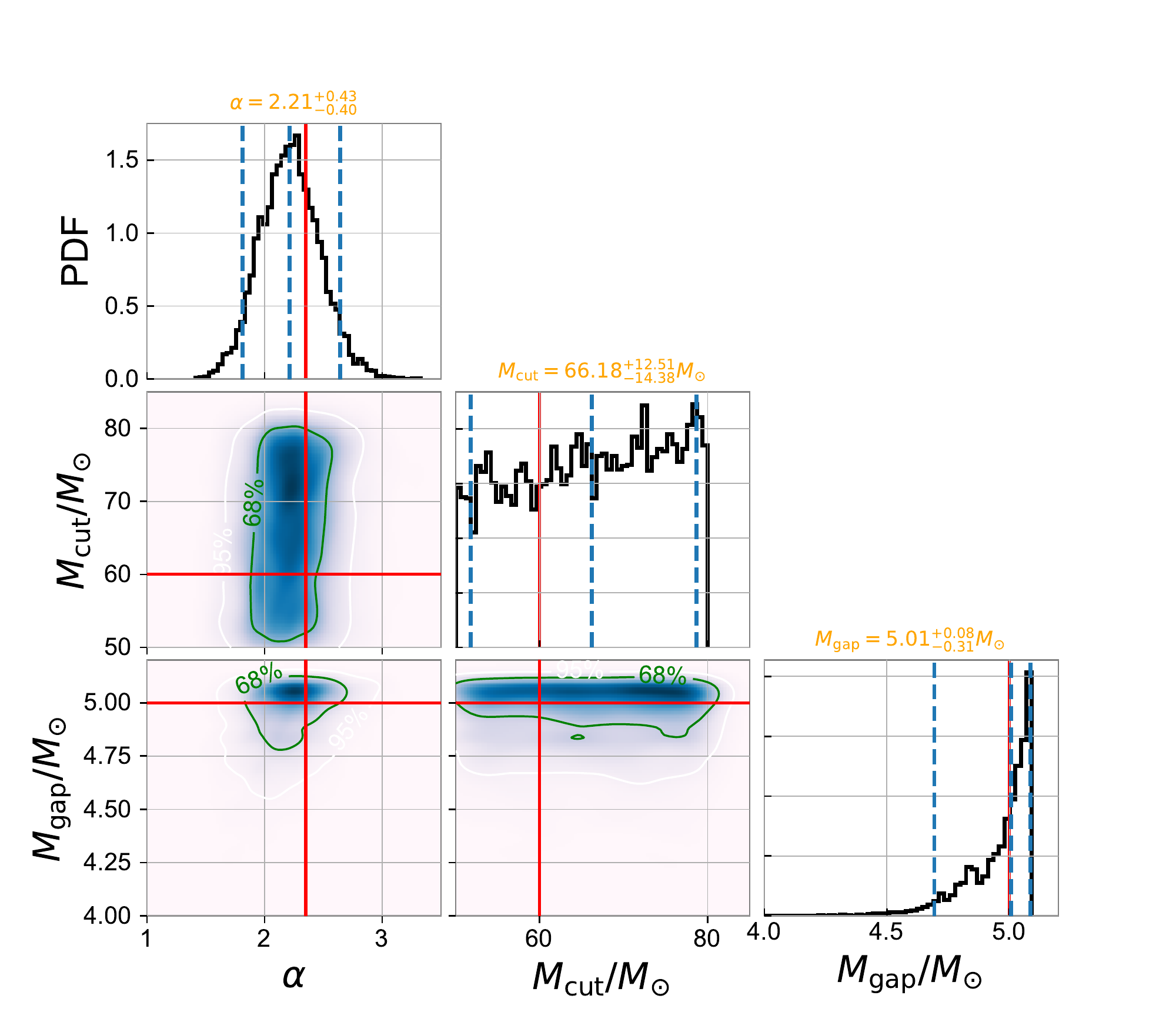}
\includegraphics[width=0.8\columnwidth]{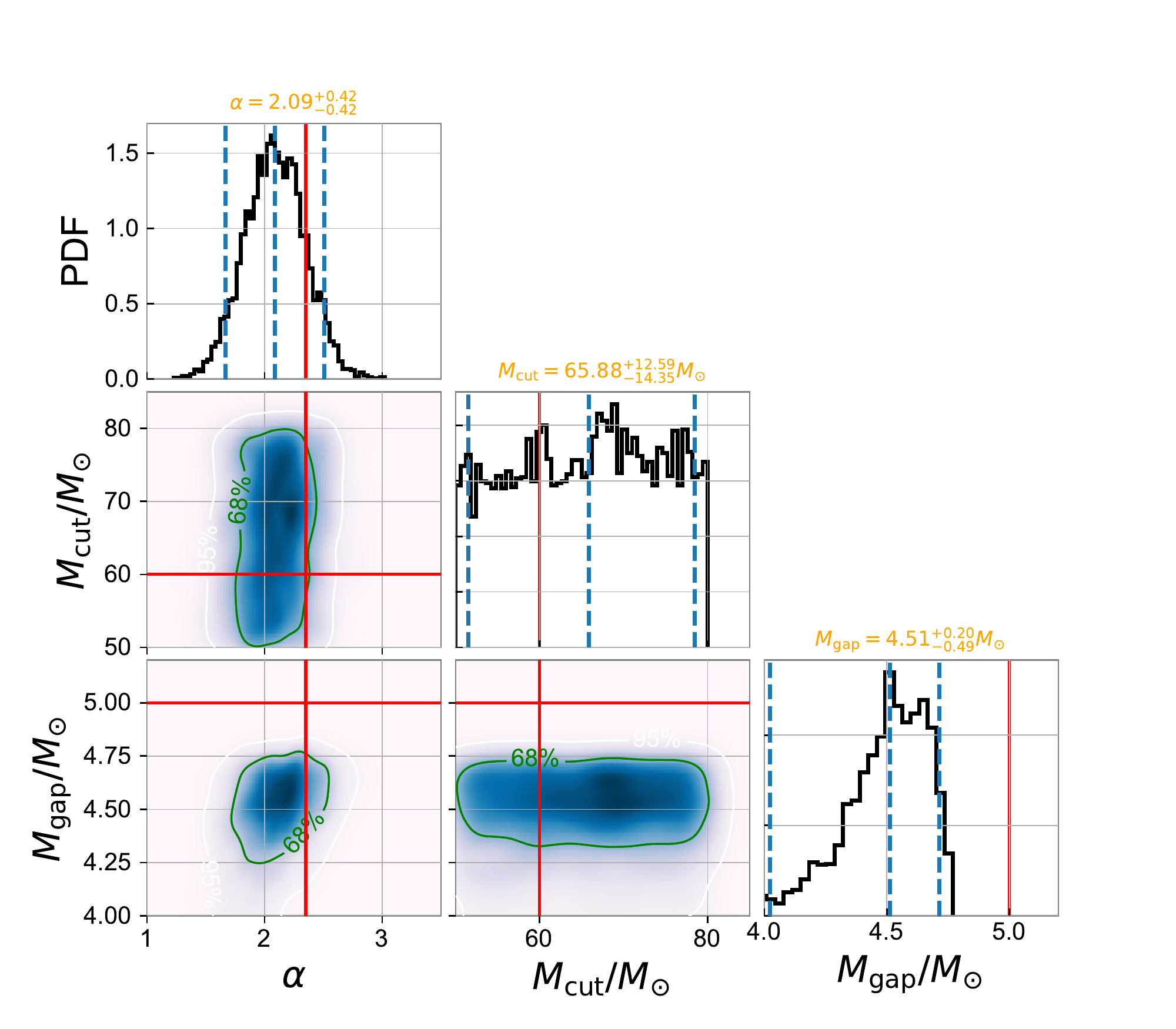}
\includegraphics[width=0.8\columnwidth]{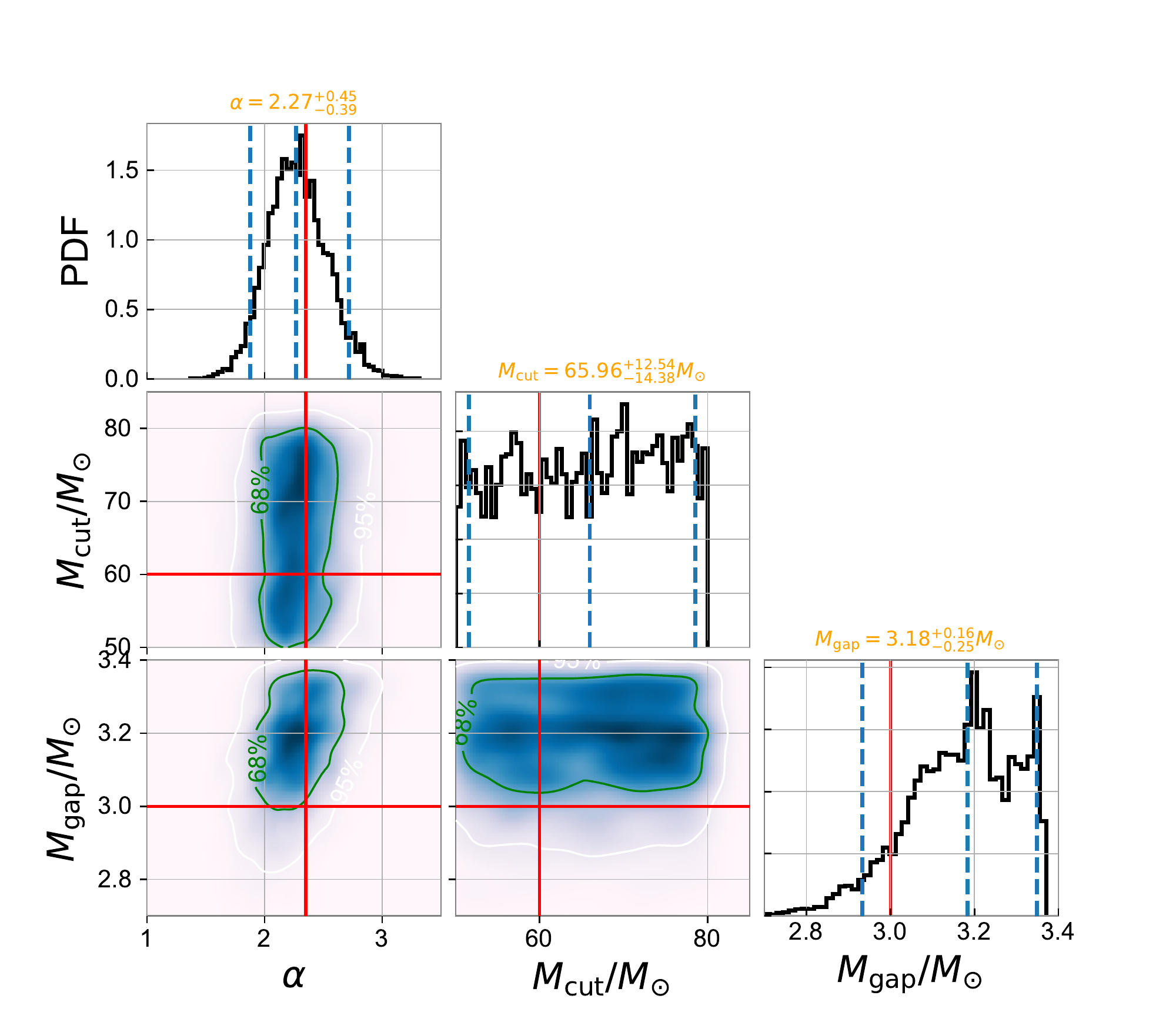}
\includegraphics[width=0.8\columnwidth]{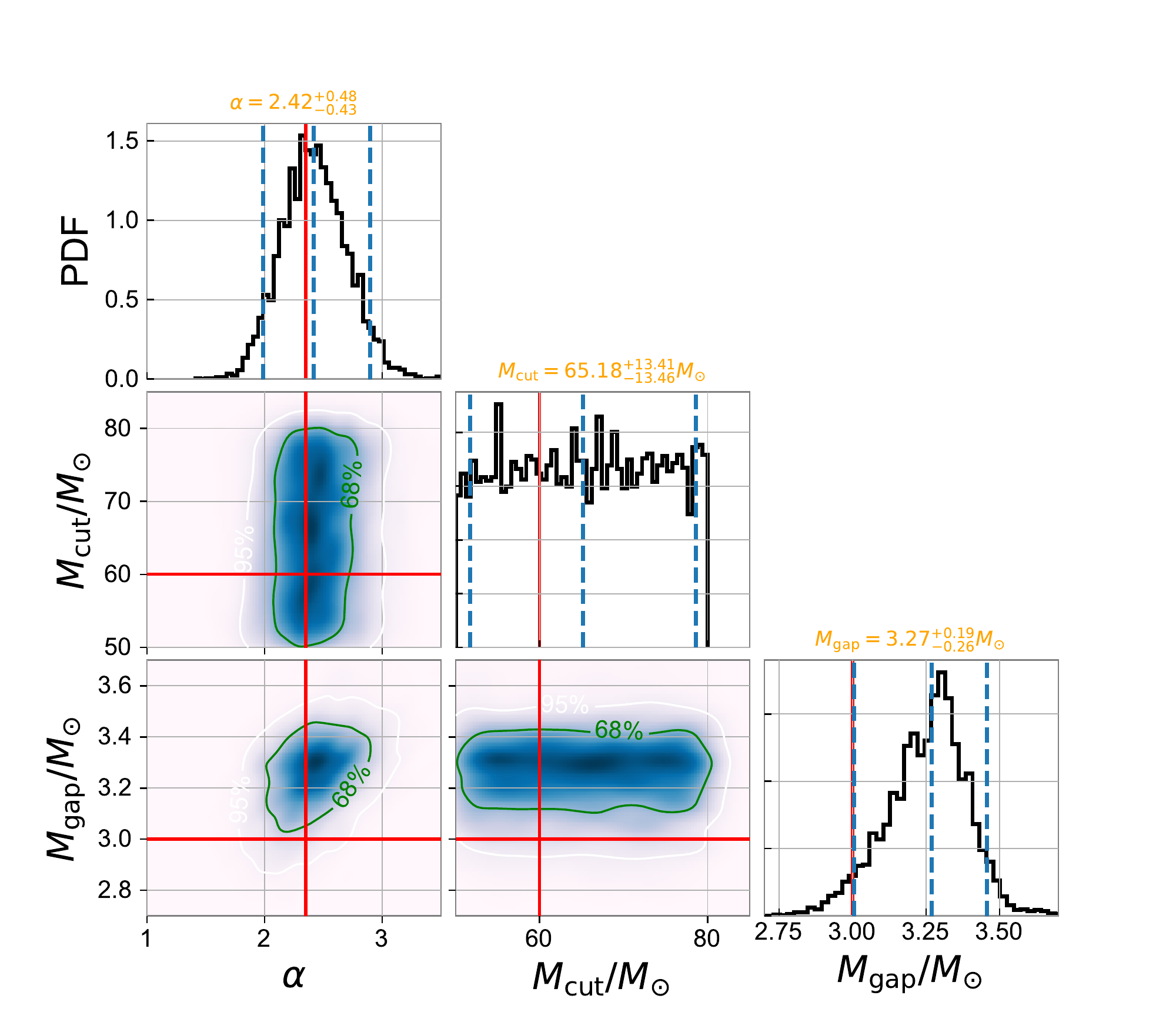}
\caption{Distributions of hyperparameters reconstructed using 50 simulated events. The top left, top right, bottom left, and bottom right panels show the results of Cases A, B, C, and D, respectively. Red lines are our fiducial values, while the dashed blue lines and the confidence intervals represent ($5\%$, $50\%$, and $95\%$) percentiles.}
\label{fig:single_bhm}
\hfill
\end{figure*}

However, due to the fact that higher mass mergers are relatively easier to detect than lower mass mergers \citep{2017ApJ...851L..25F}, we must take the selection effects into account. With the works of \citet{2019ApJ...882L..24A} and \citet{2019PASA...36...10T}, and assuming a uniform-in-log prior of rate, we can marginalize over the Poisson-distributed rate to produce the detection probability $p_{\rm det}(\Lambda | N)$ as
\begin{equation}
\label{eq:pdet}
p_{\rm det}(\Lambda | N) \propto \left(\frac{{\cal V}(\Lambda)}{{\cal V}_{\rm tot}}\right)^N,
\end{equation}
where ${\cal V}(\Lambda)$ means the ``visible volume" which can be numerically calculated with injected signals, and ${\cal V}_{\rm tot}$ refers to the total spacetime volume. The ratio of detection ${\cal V}(\Lambda)/{\cal V}_{\rm tot}$ is mainly determined by hyperparameter $\alpha$, because the power-law index describes the profile of the population properties. We simulate thousands of events with the Monte Carlo method, and collect the events above the threshold (network ${\rm S/N}>12$) of ``detecting'' GW to approximately evaluate the detection ratio (the relation between this ratio and $\alpha$ is presented in Fig.\ref{fig:pdet}). Thus the likelihood Eq.(\ref{eq:bhm}) is modified to
\begin{equation}\label{eq:selectbhm}
{\cal L}_{\rm tot}(\vec{d}, N | \Lambda, det) = \frac{1}{p_{\rm det}(\Lambda | N)}{\cal L}_{\rm tot}(\vec{d}, N | \Lambda).
\end{equation}
Finally, we take priors of the hyperparameters of BHMF as $\alpha \sim U(0.5, 5)$, $M_{\rm cut} \sim U(50, 80)\, M_{\odot}$, and $M_{\rm gap} \sim U(2.5, 6.5)\, M_{\odot}$.

\section{Results} \label{sec:results}
\begin{figure*}
\centering
\includegraphics[width=0.72\columnwidth]{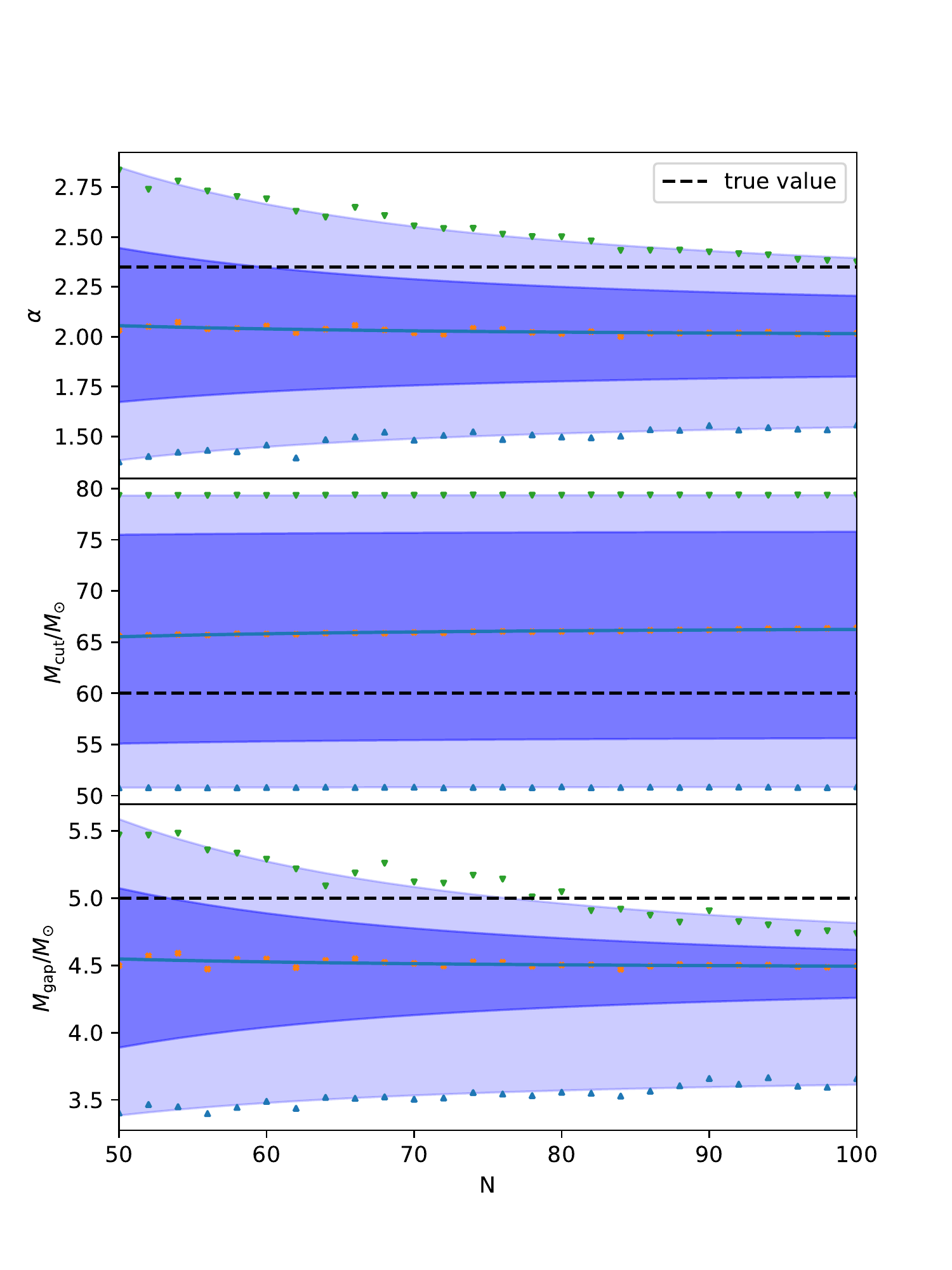}
\includegraphics[width=0.72\columnwidth]{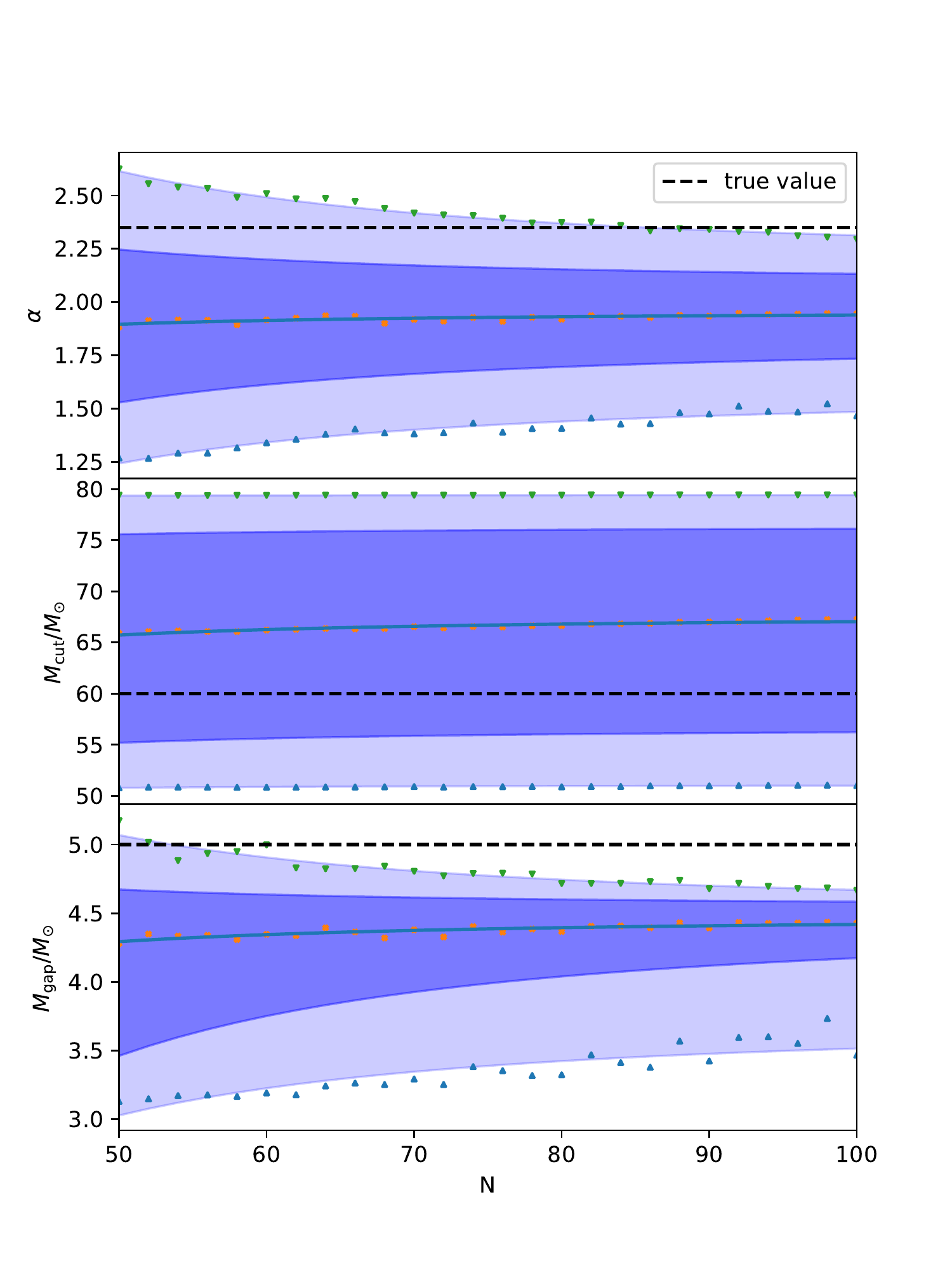}
\includegraphics[width=0.72\columnwidth]{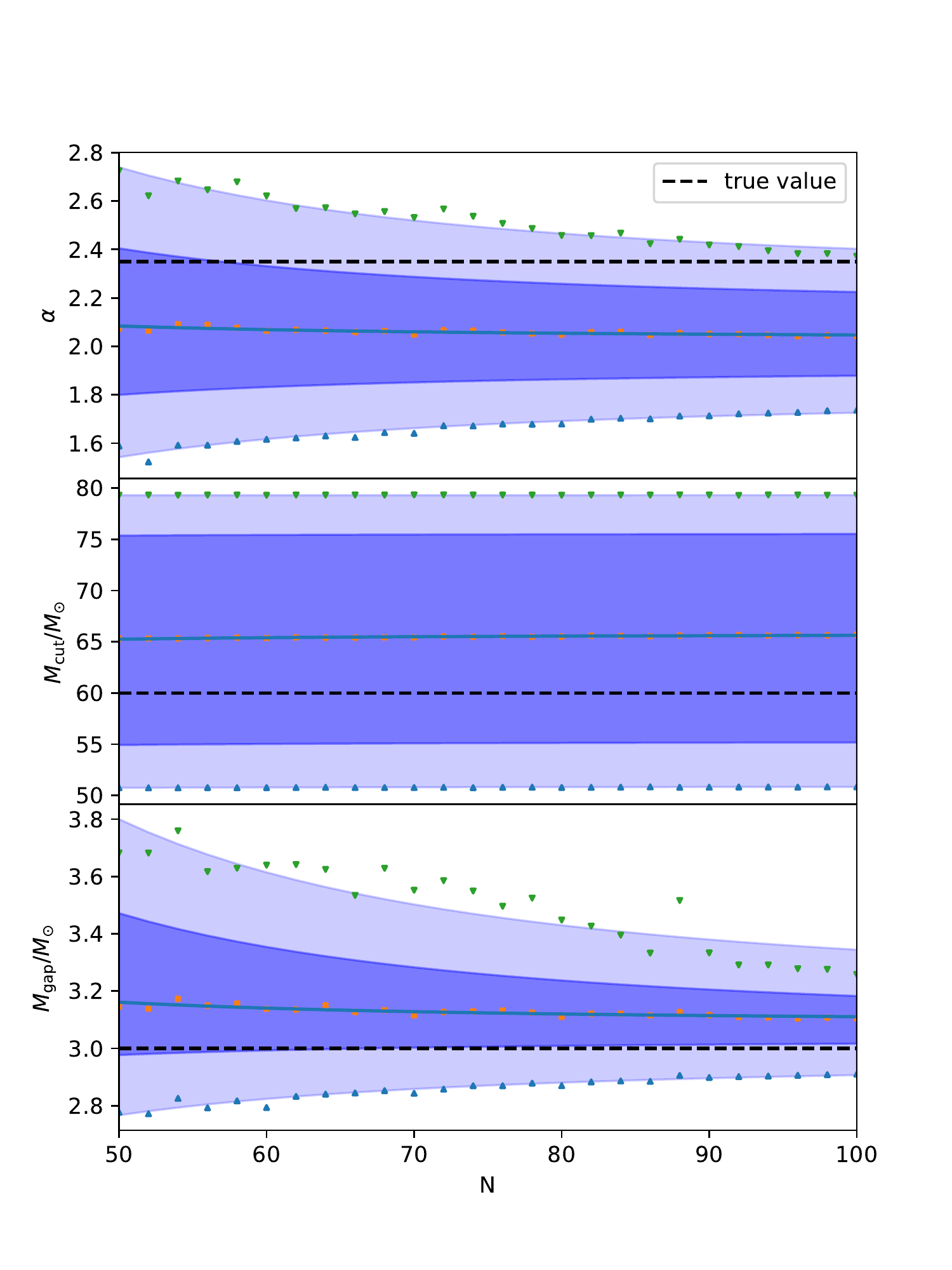}
\includegraphics[width=0.72\columnwidth]{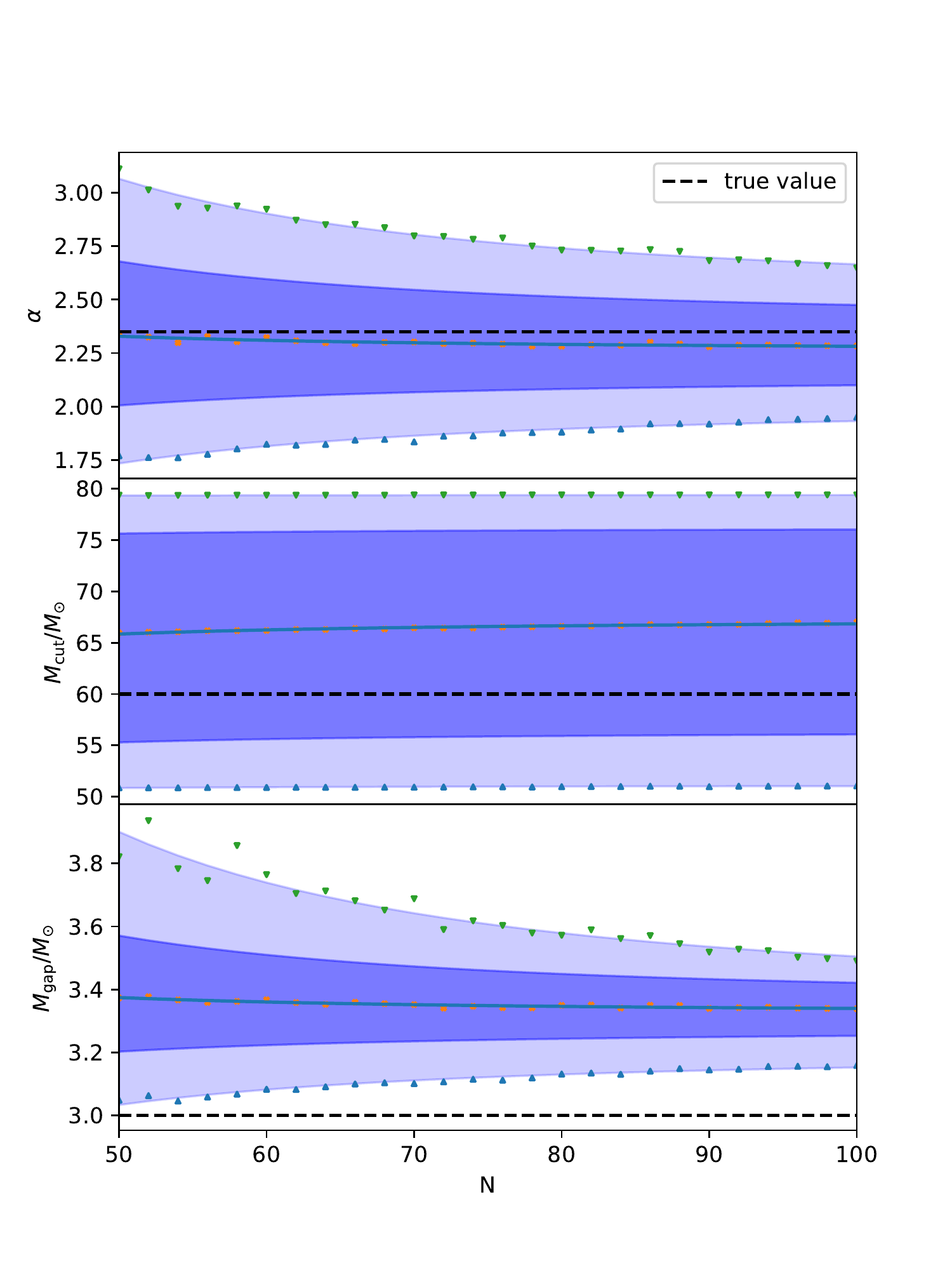}
\caption{Results of population properties of BH masses obtained with different numbers of simulated events. The top left, top right, bottom left, and bottom right panels show the confidence regions of hyperparameters reconstructed using the inferred posteriors of N events for Cases A, B, C, and D, respectively. The edges of the deep and light blue areas are smoothed with the scatter points (best-fit values and the conservative uncertainties), representing the $1\sigma$ and $2\sigma$ regions, respectively. The dashed black lines are true values of our BHMF models.
}
\label{fig:BHM}
\hfill
\end{figure*}
Applying the Bayesian parameter estimation to each simulated event, we can obtain the posterior distributions of the intrinsic parameters, e.g., mass and spin of BH. Some inference results are shown in Fig.\ref{fig:single}, and the $90\%$ confidence intervals are summarized in Table.\ref{tb:single}. As found in \citet{2015PhRvD..91d2003V}, the slightly biased median values are owing to the degeneracy between the mass ratio $q$ and spin of BH $\chi_{z}$, whose posterior distributions present a strong correlation. The degrees of biases are dependent on the {\rm S/N} and the magnitude of the BH's spin $\chi_{z}$, usually low {\rm S/N} and high $\chi_{z}$ can lead to larger mass measurement error. Because mass ratio and spin are high order post-Newtonian (PN) parameters that have minor contributions to the gravitational waves \citep{2012PhRvD..85l3007D,2019PrPNP.10903714B}, the inference of such parameters will heavily rely on the qualities of GW data. Besides, high {\rm S/N} can also reduce the uncertainties of chirp mass ($\Delta \mathcal{M}_{\rm c}/\mathcal{M}_{\rm c} \propto \mathcal{M}_{\rm c}/{\rm S/N}$; \citealt{1994PhRvD..49.2658C}), while the deviation of $\mathcal{M}_{\rm c}^{\rm src}$ is usually caused by biases of estimating luminosity distance $d_{\rm L}$ and inclination angle $\theta_{\rm JN}$. With the upgrade of Advanced LIGO/Virgo detectors, we expect to detect more and more high {\rm S/N} events. If the spins of BHs in NSBH systems share the similar properties with BBHs (i.e., Low and Restricted cases), which is beneficial for parameter estimation, then we can reduce the errors or biases of mass measurements to a certainly low level.

Therefore, it is feasible to perform a Bayesian hierarchical inference to investigate the population properties of BH masses. Though the NSBH merger rate is quite uncertain \citep{2010CQGra..27q3001A,2017ApJ...844L..22L}, four NSBH merger candidate events (S190814bv, S190910d, S190923y, and S190930t) have been claimed in public alerts of the first half-year LIGO/Virgo O3 run. The improved sensitivity in the O4 and full sensitivity runs will further enhance the detection rate significantly. Then it is reasonable to assume a sample of $\sim 50-100$ events in the next decade. Fig.\ref{fig:single_bhm} shows the results of the hyperparameters reconstructed using 50 events (randomly taken from 200 simulated events). We note that the high mass cutoff cannot be well identified in NSBH binaries, due to the very low expected number of events with $M_{\rm BH}>M_{\rm cut}$. While the gap between NS and BH is mainly determined by the event with the smallest BH mass and can be well constrained. The power-law index $\alpha$ also lies in a relatively narrow region compared to that in BBH systems \citep{2019ApJ...882L..24A}.

\begin{figure*}
\centering
\subfigure{\label{fig:class}
\includegraphics[width=1.0\columnwidth]{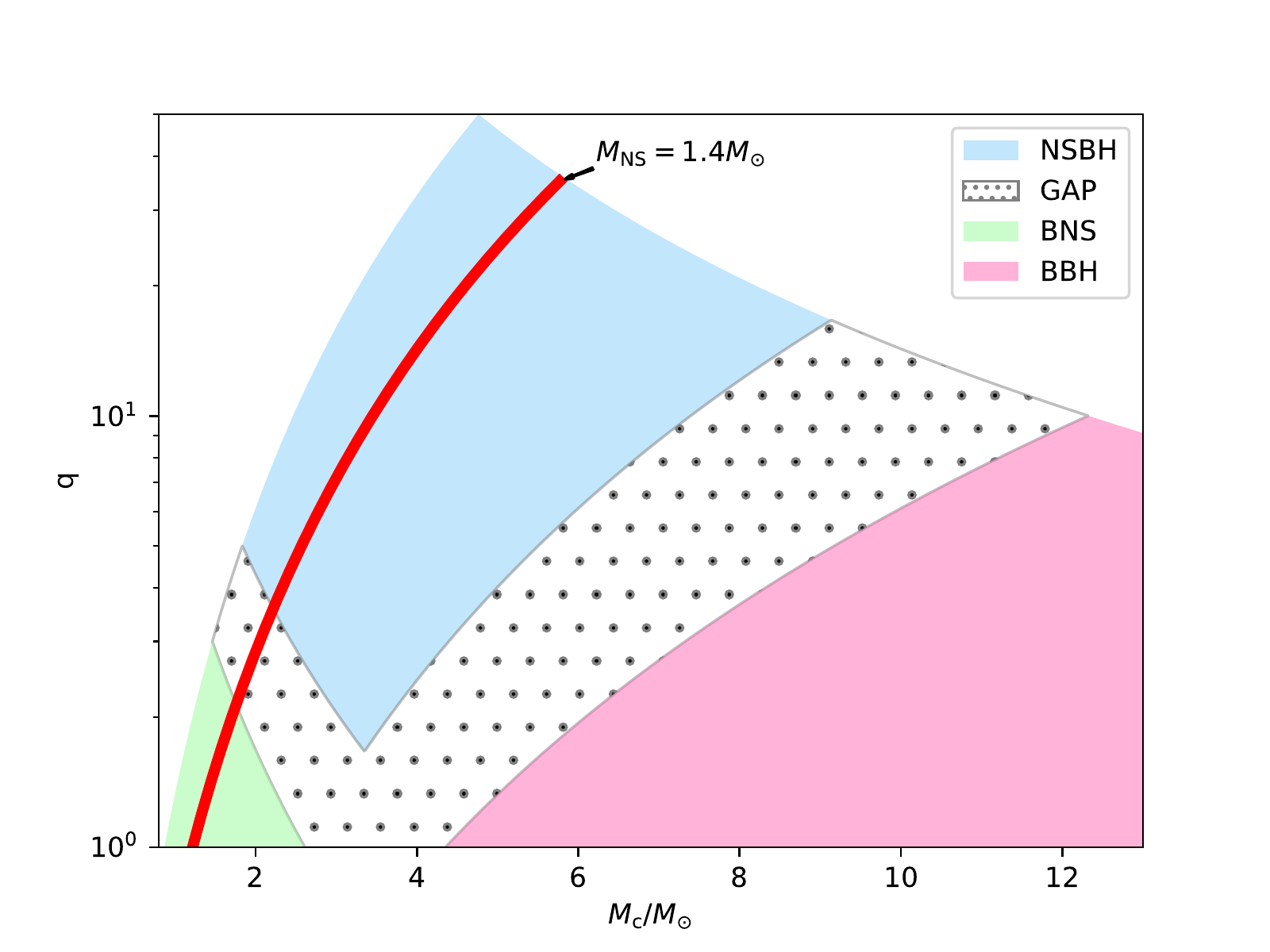}}
\subfigure{\label{fig:miss}
\includegraphics[width=1.0\columnwidth]{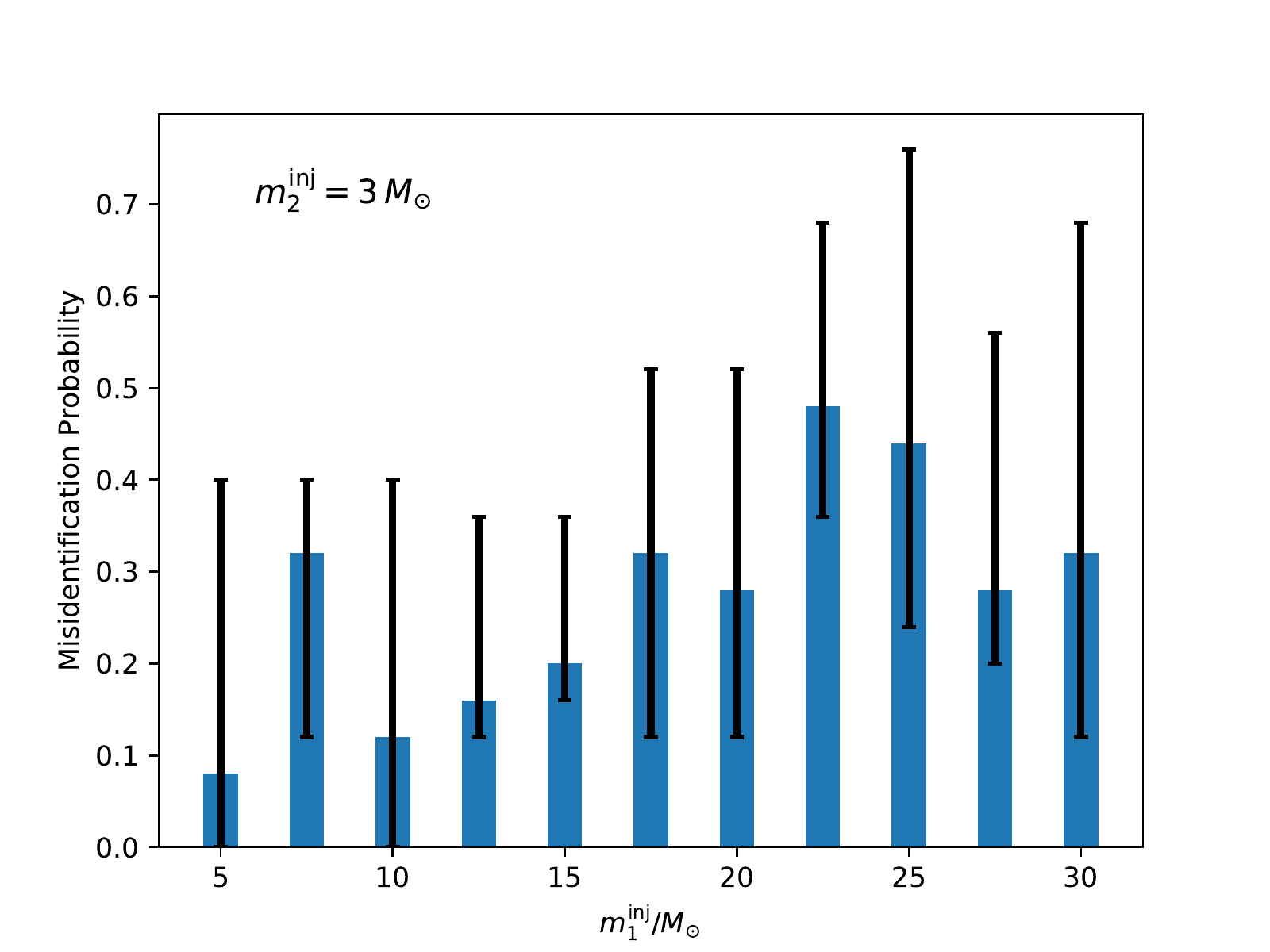}}
\caption{Left panel: classification of four mass classes (BNS, NSBH, BBH, and MassGap). Right panel: the probability of identifying BBH events as NSBH, calculated using the recovered light component masses of a series of simulated BBH merger events. The histogram and errorbars represent the ratios of simulated events with inferred masses $m_2^{\tiny{50\%}}$, $m_2^{\tiny{16\%}}$, $m_2^{\tiny{84\%}}$ (median and $1\sigma$ percentiles of posteriors of $m_2$) that are less than $3\,M_{\odot}$.}
\hfill
\end{figure*}

To make a robust evaluation of the uncertainties, we use the bootstrap method that randomly takes \emph{N} events repeating 200 times to get $\{M_{N}^1, M_{N}^2, \dots, M_{N}^{i}, \dots, M_{N}^{200}\}$. For each subset $M_{N}^{i}$ including \emph{N} groups of inferred posteriors of the \emph{N} simulated events, we fit them with the Bayesian hierarchical model using Nest sampling, and obtain the best-fit values of the hyperparameters together with their fit uncertainties $\sigma_{\rm fit}^{i}$ and the statistical uncertainties $\sigma_{\rm stat}$ (among the best-fit values). Then, we choose $\sigma=\sqrt{\overline{\sigma}_{\rm fit}^2+\sigma_{\rm stat}^2}$ as the conservative uncertainties, where $\overline{\sigma}_{\rm fit}$ is the mean value of $\sigma_{\rm fit}^{i}$. As shown in Fig.\ref{fig:BHM}, the statistical uncertainties caused by fluctuation have been greatly reduced, ensuring a better robustness on our results. By increasing the number of events, the uncertainties ($\sigma$) gradually reduced. Though, both $\alpha$ and $M_{\rm gap}$ still suffer from slight biases, we conclude that the mass gap would be identified in high significance if dozens of events are detected.

Note that in Fig.\ref{fig:single_bhm} and Fig.\ref{fig:BHM}, the possible contamination of the NSBH sample caused by the ``misclassification" of the BBH events (see Fig.\ref{fig:class}), due to the uncertainty of the measurement of $q$, has not been taken into account (some NSBH mergers in principle could also be misidentified as the BBH mergers. However, as long as the NS masses do follow a narrow distribution shown in eq.(\ref{eq:nsmf}), such a chance is very low and can be ignored). We have carried out some simulations and found out that if there is a significant mass gap between neutron stars and black holes (i.e., $M_{\rm gap}\sim 5M_\odot$), such a contamination can be ignored. However, in the absence of the mass gap, the contamination could be serious \citep[see also][]{2018ApJ...856..110Y} and the inferred $\alpha$ will be biased.
In Fig.\ref{fig:miss} we present the misclassification probability $P_{\rm mis}$, which is the chance to identify the BBH merger events with the light component mass $m_2^{\rm inj}=3\,M_{\odot}$ improperly as the NSBH ones.
Therefore, if in the future the absence of the low mass gap has been established in the BBH merger events, dedicated simulations with real PSDs are necessary to reliably infer $P_{\rm mis}$ as a function of $m_{1}$. Together with the well measured BHMF of the merging BBH systems, the contamination to the observed NSBH merger events can be effectively removed. With the ``cleaned" sample, the BHMF of the merging NSBH systems can be reasonably reconstructed. Such a detailed approach is of course beyond the scope of the current work. Though the measurement of $\alpha$ is more challenging, the absence of the low mass gap can be straightforwardly established because the mergers of the $\sim 3M_\odot$ BHs with the neutron stars usually are able to produce short GRBs and bright macronovae/kilonovae. Without the energetic neutrino emission from the central remnant and due to the higher amount of dynamical ejecta, the macronovae/kilonovae of NSBH mergers are expected to be different from that from BNS mergers \citep[e.g.,][]{Hotokezaka2013,2015ApJ...811L..22J,2020ApJ...889..171K}. Moreover, the accurately measured chirp mass can help to distinguish between the NSBH and BNS mergers and the corresponding uncertainty is better constrained for a relatively high $q$.

\section{Summary and Discussion} \label{sec:discussion}
In this work, we carry out simulations of NSBH mergers under four configurations of the population properties of BHs' spins and low mass breaks. In each case, we perform full Bayesian parameter estimations for all of the simulated events, and apply a Bayesian hierarchical model to reconstruct the parameters of population properties of BHs' masses, i.e., the hyperparameters $\Lambda=\{\alpha, M_{\rm gap}, M_{\rm cut}\}$. Though there are still biases of the recovered GW parameters in the analysis of some simulated events, the BHMF of all the cases are reconstructed with relatively small uncertainties. In the presence of a low mass gap (i.e., $M_{\rm gap}\approx 5M_\odot$), our results show a promising prospect of well measuring such a gap and studying the behavior of BHMF in different binary systems. Thus, characterizing BHMF in coalescing NSBH systems from GW measurements is feasible, which may shed new light on the formation or evolutionary paths of BHs. So far, it is unclear whether the BHMFs are different for the merging NSBH and BBH systems. Although the qualification of the prospect of identifying such a difference is beyond this work, our measurement errors of distribution parameters are relatively small (with fewer event numbers) compared with similar works on BBHs \citep{2016PhRvX...6d1015A, 2017PhRvD..95j3010K}. If the BHMFs are considerably different, e.g., $\Delta\alpha > 0.2\alpha$, it would be plausible to characterize such a difference. In the absence of a low mass gap (i.e., $M_{\rm gap}\sim 3M_\odot$), the reconstruction of the BHMF of merging NSBH systems is more challenging because of the (substantial) contamination of the BBH merger events. In this case, we need both the well-reconstructed BHMF for the merging BBH systems and the misclassification possibility of the BBH merger events into NSBH, which is obtainable via Monte Carlo numerical simulations, to reliably measure $\alpha$. The determination of the lightest BH mass $(\sim 3M_\odot)$, however, is very straightforward. This is because for such light BHs, the mergers with neutron stars will give rise to bright GRBs and in particular macronovae/kilonovae. Together with the gravitational wave data and the benefit of a relatively high $q$, this electromagnetic information can help us accurately infer the masses of the BHs. The improvements made by adding more detectors, such as  KAGRA and LIGO-India \citep{2017CQGra..34d4001A, 2018LRR....21....3A}, will be investigated in the further work. Even if the measurement precision of the parameters of a single event may not greatly increase, the increase of sensitive volume will lead to more detection events and then reduce the statistic errors, ensuring a more robust construction of BHMF in the future.

Finally, we would like to note some caveats of our results due to some model dependencies and uncertainties in the investigation. In the source parameter estimation, we have ignored some measurement errors that would appear in the real data analysis. One of them is the detector calibration error which creates uncertainties regarding strain's scale and phase. \citet{2016PhRvL.116x1102A} reported that such an error would greatly influence the sky localization but has little effect on mass measurement. So the exclusion of such an error does not influence our results. We also fix the PSD as a certain curve in the likelihood function, Eq.(\ref{eq:Likelihood}). In reality, PSD will slowly change with time. One needs to obtain the PSD from a piece of data that does not contain signals (at the time period near the event), and parameterize the PSD estimation uncertainty in likelihood function \citep{2015PhRvD..91d2003V}. In our simulations, such detailed consideration is not possible. We do not consider the systematic error caused by a waveform template, either. The template adopted in our work (i.e., SpinTaylorT4Fourier) is an inspiral-only waveform without the merger and ringdown phases, but \citet{2017PhRvL.119n1101A} showed that compared with numerical simulation waveforms containing the full inspiral-merger-ringdown phases, this waveform works well on parameter estimation. We thus expect that such template approximation is fairly good. The term of gravitational wave selection effect (i.e., $p_{\rm det}(\Lambda | N)$) is an approximate expression but has been proved effective in real data analysis \citep{2016PhRvL.116x1102A}. If we consider the effect of false alarm rate and calibration error, the only method of evaluating this is to perform Monte Carlo simulation. Another uncertainty is the detection rate that relies on the binary mass distribution and LIGO/Virgo/KAGRA's final sensitivity. Considering the high merger rate of BBHs \citep[e.g.,][]{2016ApJ...833L...1A} and binary neutron stars \citep[e.g.,][]{2018PhRvL.121p1101A,Jin2018}, there is no motivation to assume a very low merger rate of NSBH. Moreover, there are already four NSBH candidates claimed in the first half-year O3 run of aLIGO/AdV network. Recently, GW190425 \citep{2020arXiv200101761T} is also shown to be consistent with being an NSBH merger \citep{2020ApJ...891L...5H}. Therefore, a moderately large sample of NSBH mergers is expected to be accumulated in the near future, with which the black hole mass function can be reasonably reconstructed.

\acknowledgments
This work was supported in part by NSFC under grants of No. 11525313 (i.e., Funds for Distinguished Young Scholars) and No. 11921003, the Chinese Academy of Sciences via the Strategic Priority Research Program (Grant No. XDB23040000), and the Key Research Program of Frontier Sciences (No. QYZDJ-SSW-SYS024).

\software{Bilby \citep[version 0.5.5, ascl:1901.011, \url{https://git.ligo.org/lscsoft/bilby/}]{2019ascl.soft01011A}, PyCBC \citep[version 1.13.6, ascl:1805.030, \url{https://github.com/gwastro/pycbc}]{2018ascl.soft05030T}, PyMultiNest \citep[version 2.6, ascl:1606.005, \url{https://github.com/JohannesBuchner/PyMultiNest}]{2016ascl.soft06005B}}\\

\end{document}